\documentclass[notitlepage,12pt]{revtex4-1}

\usepackage{graphicx}
\usepackage{dcolumn}
\usepackage{bm}

\begin{document}

\preprint{}

\title{Is it possible to obtain cosmic accelerated expansion through  
energy transfer between different energy densities?}

\author{Recai Erdem}
\email{recaierdem@iyte.edu.tr}
\affiliation{Department of Physics \\
{\.{I}}zmir Institute of Technology\\
G{\"{u}}lbah{\c{c}}e, Urla 35430, {\.{I}}zmir, Turkey}

\date{\today}

\begin{abstract}
The equation of state of an energy 
density may be significantly modified by
coupling it to another energy density. 
In the light of this observation we check the
possibility of producing cosmic accelerated expansion 
in this way. In particular we consider the case
where matter is converted to radiation (or vice versa
by particle physics processes). 
We find that cosmic accelerated expansion can be obtained
in this way only if an intermediate state with 
negative equation of state forms during the conversion. 
\end{abstract}


\keywords{Accelerated cosmic expansion, Dark energy}

\maketitle

\section{Introduction}

Observations show that the universe is undergoing 
accelerated expansion at present, and many theoretical 
arguments and observational evidence 
suggest that the universe must have undergone an 
accelerated expansion period at the early times as well 
\cite{Weinberg, Amendola}. Although the standard explanations for 
these accelerated expansions are cosmological constant 
at present era and inflationary models at early 
times there are many alternative ways; for example, 
quintessence,  
$f(R)$ models \cite{Amendola}, and 
gravitational particle production 
\cite{Prigogine,Lima}. However all 
these models have some problems. There is a problem 
associated with cosmological constant called 
the cosmological constant problem, and it seems that the 
best way may be the use of some symmetry to make it 
cancel and seek another method for late time cosmic 
acceleration \cite{Erdem2}. Inflationary models usually employ at least
one new postulated scalar, and need special initial 
conditions, a similar situation (although less severe) 
is true for quintessence models. $f(R)$ type modified 
gravity models use an 
extension of general relativity, in gravitational particle 
production the energy density of the universe is 
an open thermodynamical system 
that is assumed to acquire energy from gravitational field
while the question of if the universe is a closed system
in this case is not clear enough.
Therefore it is useful to seek the possibility of additional
alternative ways for accelerated expansion.
In particular it would be desirable to have 
a model where the accelerated 
expansion is achieved with a minimal extension of 
the standard models of particle physics and cosmology. 
In this paper, in the light of the fact that
coupling an energy density to another one modifies its equation of
state \cite{DM-DE1} we seek if an energy density transfer 
due to elementary particle processes may have the 
potential of providing a source for cosmic accelerated 
expansion. 
Although the analysis in this paper, in principle, is 
applicable to all types of particle physics processes, 
we specify it to the case of conversion of heavy 
particles to light particles i.e. to the conversion of 
matter to radiation.
In fact  there must be an era of the creation of matter 
and radiation not only because the ordinary matter and 
radiation must be produced anyway but also to have a 
well defined model 
that may serve at all eras of the universe \cite{Erdem}. 
Moreover in the standard lore of cosmology the ordinary 
matter and radiation are assumed to be produced by the 
decays or the collisions of some other particles such as 
Higgs particle, curvaton etc. at early times 
\cite{Enqvist}. Particle physics processes ranging from 
high energies to atomic physics have an important role 
at present as well. Therefore the possibility 
of using just matter and radiation (as in this 
paper) interacting through the particle 
physics processes for cosmic acceleration with minimal 
need for exotic matter is interesting. The results of 
the following analysis shows that obtaining 
cosmic acceleration
through conversion of matter to
radiation (or vice versa) seems impossible except
through formation of an intermediate state 
with negative equation of state
(e.g. a QCD-like condensate formed by intermediate particles
produced in the particle physics processes). 

In this study we consider the Robertson-Walker metric
\begin{equation}
ds^2\,=\,-dt^2\,+\,a^2(t)[\frac{dr^2}{1-kr^2}+r^2
(d\theta^2+\sin^2{\theta}d\phi^2)] \label{b1}
\end{equation}
and for simplicity we take $k=0$ which is in agreement with 
observations \cite{PDG}.
For the illustration of the method we consider a simple 
case; a universe that consists of matter and 
radiation. We assume that, at some time $t_1$, the 
energy density of either of matter or radiation starts 
to be transferred to the other through some particle 
physics processes such as those given in Figure 
\ref{fig:1}. There may be two cases: \\
{\it i)} The conversion of each matter 
particle to radiation may be instantaneous e.g. as in 
the three particle (decay) process in Figure \ref{fig:1} or 
the four particle process in the figure where the 
internal line is deleted (i.e. a four particle contact 
interaction) or the four particle process in the figure 
where the process takes place in t or u channels. This option
is considered in the next section.\\
{\it ii)} There may be an on-shell intermediate particle in 
the process e.g. the four point process in the 
figure where the process takes place in s channel i.e. 
a resonance forms in the intermediate state in the 
conversion of matter to radiation. This option is considered
in the section after the next section.
We consider the two
cases mentioned above separately. 
In both cases above we take the matter to be extremely 
non-relativistic and approximate its cosmological 
evolution by that of dust while we assume that 
the matter particles have yet non-vanishing small velocities 
enabling them to participate in
particle processes such as scattering.

\section{The case of instantaneous interactions}

In this case the interaction between individual particles 
is instantaneous as in the three particle process or the 
four particle process in t or u channels in Fig. 
\ref{fig:1}. 
In this case the
total energy density of the universe may 
be written as 
\begin{eqnarray}
&&\rho(t)\,=\,\rho_m(t)\,+\,\rho_r(t)
\label{a1a}\\
&&
\rho_m(t)\,=\,\frac{C_{m}(t)}{[a(t)]^3}~,~~
\rho_r(t)\,=\,\frac{C_{r}(t)}{[a(t)]^4} \label{a1b}
\end{eqnarray}
where $a(t)$ is the scale factor (of Robertson-Walker 
metric) at time $t$, and (see Appendix A)
\begin{eqnarray}
&&C_m(t)\,=\,
\rho_{m0}\,-\,\int_{t_1}^{t}
\left(\frac{d\tilde{\rho}(t^\prime)}{dt^\prime}\right)
a^3(t^\prime)
\,dt^\prime \label{a2a} \\
&&C_r(t)\,=\,
\rho_{r0}\,+\,\int_{t_1}^{t}
\left(\frac{d\tilde{\rho}(t^\prime)}{dt^\prime}\right)
a^4(t^\prime)
\,dt^\prime \label{a2c} 
\end{eqnarray}
where $\frac{\tilde{\rho}(t)}{dt}$ is the rate of 
the energy density transfer from matter to radiation (in 
the commoving frame).
In this case the effect 
of interaction is absorbed into the change in evolution 
of the energy densities of matter and radiation through 
time dependence of $C_{m}(t)$ and $C_{r}(t)$.

We may easily see from 
(\ref{a1b}), (\ref{a2a}), (\ref{a2c}) 
that $\rho_m(t)$, 
$\rho_r(t)$ and
$\rho(t)\,=\,\rho_m(t)\,+\,\rho_r(t)$
satisfy the following equations
(see Appendix A) 
\begin{eqnarray}
&&
\dot{\rho}_m(t)\,+\,3H\,\rho_m(t)\,=\,
\frac{\dot{C}_{m}(t)}{[a(t)]^3}\,=\, 
-\frac{d\tilde{\rho}(t)}{dt}
\label{a2aar} \\
&&\dot{\rho}_r(t)\,+\,4H\,\rho_r(t)\,=\,
\frac{\dot{C}_{r}(t)}{[a(t)]^4} 
\,=\,\frac{d\tilde{\rho}(t)}{dt}
\label{a2bar}
\end{eqnarray}
\begin{equation}
\frac{\dot{C}_m(t)}{a^3}\,+\,
\frac{\dot{C}_r(t)}{a^4}\,=\,0
\label{a4b} 
\end{equation}
\begin{equation}
\dot{\rho}(t)\,+\,
3H\left[1\,+\,
\frac{1}{3}\frac{\rho_r(t)}{\rho(t)}
\right]\rho(t)\,=\,0
\label{a6aa}
\end{equation}
In other words the interaction that 
converts matter to radiation does not induce an 
additional energy density or pressure, so it does not 
affect the overall evolution rate of the universe while it
affects the evolution of each component, matter 
and radiation separately. In fact this conclusion holds 
in a more general context. Consider two energy densities,
$\rho_1$ and $\rho_2$ with equations of states $\omega_1$ 
and $\omega_2$, respectively in the absence of any interaction 
between $\rho_1$ and $\rho_2$; then allow an energy density 
transfer $Q$ between $\rho_1$ and $\rho_2$. The situation is similar
to Eqs. (\ref{a2aar}), (\ref{a2bar}), and (\ref{a6aa}) where 
$\frac{d\tilde{\rho}}{dt}$ is replaced by $Q$, namely,
\begin{eqnarray}
&&
\dot{\rho}_{1}(t)\,+\,3H(t)\left(1+\omega_{1}(t)\right)
\,\rho_{1}(t)\,=\,Q(t)
\label{a2aar1} \\
&&
\dot{\rho}_{2}(t)\,+\,3H(t)\left(1+\omega_{2}(t)\right)
\,\rho_{2}(t)\,=\,-Q(t)
\label{a2bar1}
\end{eqnarray}
\begin{eqnarray}
&&\dot{\rho}(t)\,+\,
3H\left(1\,+\,\omega(t)\right)
\rho(t)\,=\,0 
\label{a6aa0} \\
&&\mbox{where}~~\rho(t)\,=\,\rho_1(t)+\rho_2(t)~,~~
\omega(t)\,=\,\frac{
\omega_1\rho_1(t)+
\omega_2\rho_2(t)}{\rho_1(t)+\rho_2(t)} \label{a6aa00}
\end{eqnarray}
In fact the form of the equations (\ref{a2aar1}) 
and (\ref{a2bar1}) are familiar from the models of 
dark matter - dark energy interactions \cite{DM-DE} where dark matter and
dark energy are replaced by (dark) matter and radiation. 
It is evident from the form of (\ref{a6aa00}) that
$\omega$ is positive for all times if both $\omega_1$
and $\omega_2$ are positive at some initial time and 
$\omega$ is negative for all times if both $\omega_1$
and $\omega_2$ are negative at some initial time (provided that
$\rho_{1(2)}$ is positive). In other words $\omega_1$
and $\omega_2$ should have different signs if one wants to 
construct a model where the dark energy is subdominant initially and
at later time becomes dominant or vice versa (and introduction of
$Q$ does not have any effect on the time of transition between 
$\omega\,>\,0$ and 
$\omega\,<\,0$ eras). 
Therefore one can not
induce a dark energy effect through conversion of matter to radiation
(or vice versa) in this way.

For example, assume that the conversion rate of matter and
radiation is such that $-\frac{d\tilde{\rho}(t)}{dt}$=$3H\rho_m$.
Then (\ref{a2aar}), (\ref{a2bar}) become 
\begin{eqnarray}
&&\dot{\rho}_m(t)\,+\,3H\,\rho_m(t)\,=\,
3H\rho_m(t)
~~~~\Rightarrow~~
\rho_m(t)\,=\,\mbox{constant}\,=\,
\rho_0
\label{a2arr} \\
&&\dot{\rho}_r(t)\,+\,4H\,\rho_r(t)\,=\,
-3H\rho_m(t)\,=\,
-3H\rho_0 
\label{a2arrr}
\end{eqnarray}
Hence
\begin{equation}
\frac{4}{3}\rho_r+\rho_0\,=\,
A\,e^{-3\int\,H\,dt}
\label{a2barr}
\end{equation}
where $A$ is some integration constant.
In other words $\rho_m$ in this example behaves
as cosmological constant (although in the absence of
an energy density transfer between matter and radiation $\rho_m$ 
scales as $\frac{1}{a^3}$), and $\rho_r$ behaves
as an exotic fluid with the energy density given 
in (\ref{a2barr}). Therefore introduction of an energy transfer
between matter and radiation may considerably change their cosmic
evolution and an otherwise standard energy density may seem exotic
due to the coupling of that energy density to another energy density.
However if one adds the differential equations in 
(\ref{a2arr}) and
(\ref{a2arrr}) one finds that the equation is the same as 
(\ref{a6aa}) i.e. it is the same as the evolution of an energy density
composed of matter and radiation that do not interact with each other.
In other words, although the evolution of each energy density component
in total energy density may drastically change due to introduction of an energy
transfer between different components of the total energy density, the overall 
evolution of the total energy density remains the same after the introduction of 
the energy transfer.

\section{The case of a resonance as the intermediate 
state in the conversion of matter to radiation}
\subsection{General discussion}

This is the case where the internal line in the four 
particle process in Fig. \ref{fig:1} is an on-shell 
s-channel 
intermediate particle (i.e. resonance). 
In this case the
total energy density of the universe may 
be written as 
\begin{eqnarray}
&&\rho(t)\,=\,\rho_m(t)\,+\,\rho_r(t)
\,+\,\rho_R(t) 
\label{a1ar}
\end{eqnarray}
where 
$\rho_m(t)$ and 
$\rho_r(t)$ are the energy densities of matter and
radiation given by 
(\ref{a1a}) where 
$C_{m}(t)$ and 
$C_{r}(t)$ are replaced by
\begin{eqnarray}
&&C_m(t)\,=\,
\rho_{m0}\,-\,\int_{t_1}^{t}
\left(\frac{d\tilde{\rho}_m(t^\prime)}{dt^\prime}\right)
a^3(t^\prime)
\,dt^\prime \label{a2ar} \\
&&C_r(t)\,=\,
\rho_{r0}\,+\,\int_{t_1}^{t}
\left(\frac{d\tilde{\rho}_r(t^\prime)}{dt^\prime}\right)
a^4(t^\prime)
\,dt^\prime \label{a2cr} 
\end{eqnarray}
Because the form of 
the energy density of the intermediate state i.e. the resonance
$\rho_R(t)$ is not specified it has a general form
given by  
\begin{equation} 
\rho_{R}(t)\,=\,
e^{-\Gamma(t-t_1)}\,
C_R(t)
\,\exp{\left[-3\int_{a(t_1)}^{a(t)}
\frac{\left(1+\omega_R(a^\prime)\right)\,da^\prime}
{a^\prime}\right]} \label{a21a}
\end{equation}
Note that 
$e^{-\Gamma(t-t_1)}$ term would not be present in (\ref{a21a}) 
if we had taken the resonance as stable,	
where $\Gamma$ is the width of the resonance.
$C_R$ in (\ref{a21a}) may be obtained in the same way as done in
Appendix A as
\begin{equation}
C_R(t)\,=\,
\int_{t_1}^{t}dt^\prime[
(\frac{d\tilde{\rho}_m(t^\prime)}{dt^\prime})
\,e^{\Gamma(t^\prime-t_1)}\,
\exp{\left[3\int_{a(t_i)}^{a(t^\prime)}
\frac{\left(1+\omega_R(\tilde{a})\right)d\tilde{a}}
{\tilde{a}}\right]}
\label{a21}
\end{equation} 
where we have used
$\frac{d\tilde{\rho}_R(t^\prime)}{dt^\prime}$=
$\frac{d\tilde{\rho}_m(t^\prime)}{dt^\prime}$
since the amount of energy transferred from the matter
is equal to the one acquired by the resonance.
$\rho_R$ satisfies
\begin{eqnarray}
&&\dot{\rho}_R\,+\,3H(1+\omega_R)\rho_R\,=\,
\,e^{-\Gamma(t-t_1)}\,
\dot{C}_{R}(t)
\,\exp{\left[-3\int_{a(t_i)}^{a(t)}
\frac{\left(1+\omega(a^\prime)\right)}
{a^\prime}\right]}\nonumber\\
&&-\Gamma\,
\,e^{-\Gamma(t-t_1)}\,
C_{R}(t)
\,\exp{\left[-3\int_{a(t_i)}^{a(t)}
\frac{\left(1+\omega(a^\prime)\right)}
{a^\prime}\right]}
\,=\,\frac{d\tilde{\rho}_R(t)}{dt}
\,-\,\Gamma\,\rho_R
\label{a21b}
\end{eqnarray}
which is equivalent to
\begin{equation}
\dot{\rho}_R\,+\,3H(1+\omega_R+
\Delta\omega_R)\rho_R\,=\,o
\label{a21c}
\end{equation}
where
\begin{equation}
\Delta\omega_R\,=\,\frac{1}{3H}\left(
-\frac{\frac{d\tilde{\rho}_R(t)}{dt}}
{\rho_R}\,+\,\Gamma\right) \label{a21d}
\end{equation}
Eq.(\ref{a21d}) tells us that the 
effect induced by the formation of the resonance
will act like dark energy if the ratio of the rate of 
conversion of the energy (of the matter to the resonance) 
to the energy density of the resonance
is higher than the decay
rate of the resonance. Note that this effect is independent
of the background effect produced by the resonance i.e.
$\Delta\omega_R$ is independent of $\omega_R$ in general.

Note that we had taken 
$\frac{d\tilde{\rho}_m}{dt}$=
$\frac{d\tilde{\rho}_r}{dt}$ in the previous case because the 
conversion of each matter particle to radiation was 
instantaneous while here 
$\frac{d\tilde{\rho}_m}{dt}\,\neq\,
\frac{d\tilde{\rho}_r}{dt}$ since there is a resonance as the 
intermediate state in this case hence the rate of 
the conversion of matter to the resonance, 
$\frac{d\tilde{\rho}_m}{dt}$ may not be equal to 
the rate of the conversion of the energy density of the 
resonance to radiation 
$\frac{d\tilde{\rho}_r}{dt}$ 
at a specific time. Instead we should
have
\begin{eqnarray}
&&-\frac{d\tilde{\rho}_m}{dt}\,+\,
\frac{d\tilde{\rho}_r}{dt}\,+\,
\frac{d\tilde{\rho}_m}{dt}\,-\,
\Gamma\,\rho_R \nonumber \\
&&=\,
\frac{\dot{C}_m(t)}{a^3}\,+\,
\frac{\dot{C}_r(t)}{a^4}\,+\,
\left[\frac{d\left(
e^{-\Gamma(t-t_1)}\,
C_R(t)\right)}{dt}\right]\,\exp{\left[-3\int_{a(t_1)}^{a(t)}
\frac{\left(1+\omega(a^\prime)\right)}
{a^\prime}\right]}
\,=\,0
\label{r1a} \\
&&\mbox{i.e.}~~~~
\frac{d\tilde{\rho}_r}{dt}\,=\,
\Gamma\,\rho_R
\label{r1b} 
\end{eqnarray}
In other words the net change in the total energy due to the 
energy transfers between matter, radiation, and the resonance
state should be zero at each time $t$ since we assume that
system is a closed system with no energy transfer from outside, 
and the source of radiation is the decay of the resonance.
Eq.(\ref{r1a}) tells us that 
there is no direct effect of the interaction (of the
conversion of energy densities) on the total energy density,
instead the effect of conversion is indirect and through
the evolutions of $\dot{C}_m(t)$, $\dot{C}_r(t)$, $\dot{C}_R(t)$
as in the preceding subsection. In other words by taking different
sets of $\dot{C}_m(t)$, $\dot{C}_r(t)$, $\dot{C}_R(t)$ one may
mimic different effective energy densities in this case as well while
the evolution of the total energy density does not change after 
introduction of conversion between energy densities of matter and radiation.
The only way to induce a dark energy effect in this way is to assume that the
resonance state has $\omega\,<\,0$ (e.g. the resonance particles form a QCD-like
condensate). In fact this may be anticipated from (\ref{a6aa00}); one may introduce
a third energy density $\rho_3$ with $\omega_3\,<\,0$ (or a variable $\omega_3$ that 
becomes $\omega_3\,<\,0$ at some time during its evolution) so that the equation
state of the universe, $\omega$ may change sign during the evolution of the
energy densities while it is impossible in the case of a universe that only
consist of matter and radiation with 
$\omega_1\,>\,0$, $\omega_2\,>\,0$, $\rho_1\,>\,0$, $\rho_2\,>\,0$.
In this subsection we want to focus on the effect of the resonance and
possibility of taking it as an energy density inducing an accelerated 
cosmic expansion.

One may describe the general evolution of this system as follows:
The evolutions of $C_m$ and $C_R$ are given by  (\ref{a2ar}) and
(\ref{a21}), respectively. The evolution of $C_r$ is given by 
(\ref{a2cr}) where (\ref{r1b}) is employed.
For simplicity we take $\rho_{r0}=0$ i.e.
\begin{equation}
C_r(t)\,=\,
\Gamma\,\int_{t_1}^{t}
\rho_R(t^\prime)
a^4(t^\prime)
\,dt^\prime \label{a2cr1}
\end{equation}
Then at $t=t_1$ $C_m$ (so, $\rho_m$) is maximum, and 
$C_R$ (so, $\rho_R$) and $C_r$ (so, $\rho_r$) are zero.
By time $C_m$ decreases till $t=t_{mf}$ when $C_m$ becomes 
small or $C_m\simeq\,0$ 
and $C_R$ is maximum and $C_r$ has a small finite value.
At $t\,>\,t_{mf}$ $C_m$ gets even closer to zero (if it is not zero yet), 
$e^{-\Gamma\,(t-t_1)}\,C_R(t)$ gets smaller by time and 
becomes zero as $t\rightarrow\,\infty$ while $C_r$ gets its 
maximum at $t\rightarrow\,\infty$. In other words the evolution
of the system may be summarized as: Initially the universe is
matter dominated then it becomes resonance dominated and 
eventually the universe becomes radiation dominated. 
In the next subsection
we will describe the resonance state by
resonance particles produced by the particle physics process
given in the second diagram in Fig.\ref{fig:1} for relatively
simple cases.

\subsection{A particle physics description}

We may adopt the derivation for 
the formula of energy density of matter in 
Appendix A (i.e. Eq.(\ref{a2ar})) to find an expression for number 
densities since both have the same dependence on scale factor 
i.e. both vary as $\propto\,a^{-3}$. To find the corresponding 
expression for number densities we replace 
$\frac{d\tilde{\rho}(t)}{dt}$ by
$\frac{d\tilde{n}(t)}{dt}$ and $\rho_m(t)$ by $n(t)$.
Then we may employ the method we had used for energy densities 
in Appendix A to write a similar expression for number
densities of stable particles
\begin{equation}
n(t)\,=\,
\frac{C^{(n)}(t)}{[a(t)]^3}~,~~~~
C^{(n)}\,=\,
n_1\,-\,\int_0^{t-t_1}
\left(\frac{d\tilde{n}(x)}{dx}\right)_{x=(t-u)}a^3(t-u)
\theta(u)\,du \label{ca}
\end{equation}
where we have put the superscript $^{(n)}$ to the coefficient 
$C$ to distinguish it from the corresponding symbol for energy 
densities, $n_1$ stands for the initial value of $n(t)$ at
the initial time $t=t_1$ of the energy transfer, and 
$\frac{d\tilde{n}(t)}{dt}$
stands for the number density transferred per time in 
per comoving volume i.e. the rate of the number density 
transfer in the comoving frame, and $\theta(u)$ is the Heaviside function
(i.e. unit step function) given by 
$\theta=0$ if $u\,<\,0$ and
$\theta=1$ if $u\,>\,0$. 
Eq.(\ref{ca}) 
(after using a derivation similar to the one
for matter in Appendix A) 
implies that
\begin{equation}
\frac{d\tilde{n}(t)}{dt}
\,=\,
-\frac{\dot{C}^{(n)}(t)}{[a(t)]^3}
\label{ca1}
\end{equation}

\begin{figure}[h]
\begin{picture}(200,150)(50,50)
\put (0,0){\line(1,0){40}}
\put(10,10){$p_1$}
\put (40,0){\vector(1,1){25}}
\put (65,25){\line(1,1){25}}
\put(90,40){$p_2$}
\put (40,0){\vector(1,-1){25}}
\put (65,-25){\line(1,-1){25}}
\put(90,-30){$p_3$}
\hspace*{70pt}
\put (105,50){\vector(1,-1){25}}
\put (130,25){\line(1,-1){25}}
\put(108,20){$p_1$}
\put (105,-50){\vector(1,1){25}}
\put (130,-25){\line(1,1){25}}
\put(108,-25){$p_2$}
\put (155,0){\line(1,0){25}}
\put (180,0){\vector(1,1){25}}
\put (205,25){\line(1,1){25}}
\put(220,30){$p_3$}
\put (180,0){\vector(1,-1){25}}
\put (205,-25){\line(1,-1){25}}
\put(220,-30){$p_4$}
\end{picture}
\hspace{10pt}\\
\hspace{10pt}\\
\hspace{10pt}\\
\hspace{10pt}\\
\hspace{10pt}\\
\hspace{10pt}\\
\caption{The diagram on the left-hand side shows 
the decay of a particle with momentum $p_1$ into 
two particles with momenta $p_2$ and $p_3$ e.g. the 
decay of a non-relativistic particle to two relativistic 
particles while the diagram on the right-hand side 
shows the (inelastic) collision of two particles 
with momenta $p_1$ and $p_2$ into 
two other particles 
with momenta $p_3$ and $p_4$ 
e.g. the collision of two 
non-relativistic particles into two relativistic 
particles through formation of an intermediate state} 
\label{fig:1}
\end{figure}
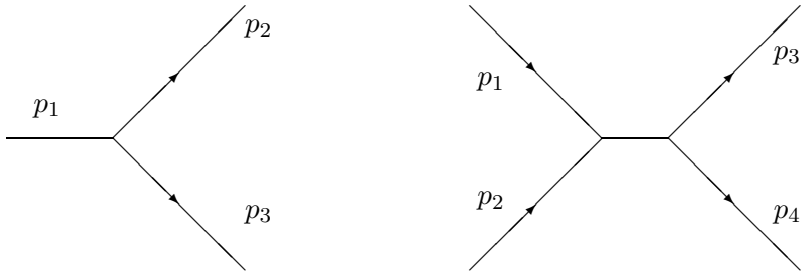

The rate of the conversion of the matter to resonance
particles by collisions of two beams of
matter particles, each with number densities 
$\frac{n_m}{2}$ may be expressed as
\begin{equation}
-\frac{d\tilde{n}_m(t)}{dt}
\,=\,
\frac{\dot{C}^{(n)}_m(t)}{[a(t)]^3}
\,=\,
-\beta^\prime\,n_m^2(t)\,\sigma(t)\,v(t)
\label{c1}
\end{equation}
where $\frac{1}{4}\,>\,\beta^\prime\,>\,0$ is some constant 
designating the effective penetration depth of a beam
of $\frac{n_m}{2}$ to a target of 
$\frac{n_m}{2}$ particles in unit comoving volume, 
$\sigma$ is the cross section of
the four-point Feynman diagram in Fig.\ref{fig:1}, 
$v$ is the average
velocity of the (non-relativistic) particles relative
to the comoving frame. For later reference we need 
dependence of cross section on scale factor. The differential
cross section $d\sigma$ of two initial particles of 4-momenta
$p_1$, $p_2$ and masses $m_1$, $m_2$ 
going to a final state of two particles of 4-momenta 
$p_1^\prime$ and $p_2^\prime$
in local Minkowski space 
is given by \cite{Langacker}
\begin{eqnarray}
d\sigma\,=\,\frac{(2\pi)^4\delta^4(p_1+p_2-p_1^\prime-p_2^\prime)}{4E_1E_2v}
\frac{d^3{\bf p}_1^\prime}{(2\pi)^32E_1^\prime}
\frac{d^3{\bf p}_2^\prime}{(2\pi)^32E_2^\prime}
|M_{fi}|^2
\label{cr1}
\end{eqnarray}
where bold face quantities denote 3-dimensional quantities, $E$ denotes energies, 
the subindices $1$ and $2$ denote the 1st and the 2nd particles, unprimed
letters correspond to initial state particles and the primed ones correspond to final state particles.
$E_j$ denotes energy of the j'th particle, and $M_{fi}$ is the transition amplitude from
initial state to the final state. The dependence of the cross section
on scale factor is through its dependence on momenta in (\ref{cr1}). After integrating
over final momenta the apparent dependence of the cross section on ${\bf p}_{1(2)}^\prime$ disappears.
Therefore the dependence of the cross section on scale factor is through 
$E_{1(2)}$, $E_{1(2)}^\prime$, and $|M(p_1,p_2;k_i)|^2$. In this case 
$E_{1(2)}$ are extremely non-relativistic, so they may essentially be taken constant.
$E_{1(2)}^\prime$ are extremely relativistic, so may be taken to vary proportional to 
$\frac{1}{a}$. $M_{if}$ corresponding to the four-point process in figure in the perturbative
limit for an on-shell intermediate state (in s-channel) is given by \cite{Langacker,comp-mass}
\begin{eqnarray}
M_{fi}\,=\,(-ig_1})({-ig_2)\left(\frac{1}{s-m_R^2+i\sqrt{s}\Gamma_R}\right)
\label{cr2}
\end{eqnarray} 
where $g_1$, $g_2$ are coupling constants of the interactions in the vertices,
$s=(p_1+p_2)^2$, and $m_R$, $\Gamma_R$ are the mass, the 
decay width of the resonance, respectively. Because the initial state consist of
extremely non-relativistic particles one may write
\begin{equation}
(p_1+p_2)^2\,\simeq\,(E_1+E_2)^2\simeq\,(2m+mv^2)^2\;,~~m_R\simeq\,2m \nonumber 
\end{equation}
Therefore for the values of $s$ near the pole of the resonance 
\begin{equation}
s-m_R^2
+i\sqrt{s}\Gamma_R(s)\,\simeq\,m_R^2v^2
+im_R\Gamma_R\left(s\right)
\label{cr3}
\end{equation}
$\Gamma_R$ is also a function of $s$ in general, as emphasized in (\ref{cr3}), 
so the form of dependence of $\Gamma_R$ on $s$ must be also known to 
determine the dependence of cross section on scale factor. In the case 
of perturbative field theory $\Gamma_R$ is identified as the imaginary
part of self interaction and can be calculated in principle. In the case of
a perturbative quantum field theory consisting of only scalars with 3-point
interactions the leading order contribution is found to be proportional to 
$\sqrt{s}$ \cite{self-energy}. When the theory gets non-perturbative higher order terms
that depend on higher powers $a$ are expected to become dominant. 
In fact in this study we consider extremely narrow resonances (i.e. $\Gamma\,<<\,m_R$) 
since the lifetime of
the resonance must be long enough to be cosmologically relevant, so
the last term in (\ref{cr3}) may be neglected.
Therefore
\begin{equation}
M_{fi}\,\propto\,a^p(t)
\label{cr4}
\end{equation}
where $p=2$ in the case of an extremely narrow resonance.
Hence we find that
\begin{equation}
d\sigma\,\propto\,a^r(t)~~~~\mbox{where in the case of an extremely narrow resonance}~~
~r\,=\,7 \label{cr5}
\end{equation}
The same result may be obtained from the general expression
$M_{fi}$ for the values of
the total energy close to the resonance pole \cite{Salam,PDG}
\begin{equation}
M_{fi}\,=\,-\Gamma_1\Gamma_2\left(\frac{1}{s-m_R^2+i\sqrt{s}\Gamma_R}\right)
\label{cr6}
\end{equation} 
where $\Gamma_{1(2)}$ are the partial widths of the resonance associated with
the first and second vertex in Figure \ref{fig:1}. While (\ref{cr6}) is more
general in the sense that it is not restricted to perturbative interactions, 
it may not be evident that $\Gamma_{1(2)}$ do not vary with scale factor.

The energy density of dust particles 
is proportional to its number density. We assume
the matter particles are extremely 
non-relativistic so their energy density may be approximated
by that of dust. Therefore (\ref{c1}) may be rewritten as
\begin{equation}
\frac{d\tilde{\rho}_m(t)}{dt}
\,=\,E
\frac{d\tilde{n}_m(t)}{dt}
\,=\,
-\frac{\dot{C}_m(t)}{[a(t)]^3}
\,=\,
\beta\,\rho_m^2(t)\,a^{r-1}(t)~~,~~
\beta\,=\,
\frac{\beta^\prime\,\sigma(t_1)\,v(t_1)}{E\,a(t_1)}
\label{c1a1}
\end{equation}
where $E\simeq\,m$ is the average energy of the matter
particles.
In (\ref{c1a1}) we have used Eq.(\ref{cr4})
i.e. $\sigma(t)$=$\sigma(t_1)
\left(\frac{a(t)}{a(t_1)}\right)^r$ (where
we have used the expression for cross section
in local Minkowski space), and the fact that 
$v(t)$=$v(t_1)\frac{a(t_1)}{a(t)}$, and $E$ 
is taken to be constant since particles are taken to 
be extremely non-relativistic and it becomes
even more non-relativistic by time due to redshift. 
After using $\rho_m(t)$ in (\ref{a1b}), (\ref{c1a1}) may 
be rearranged as 
\begin{eqnarray}
&&\frac{\dot{C}_m(t)}{[a(t)]^3}
\,=\,
-\theta(t-t_1)\,
\beta\,
\left(\frac{C_m(t)}{[a(t)]^3}\right)^2\,
a^{r-1}(t)
\label{ca2}\\
&&\Rightarrow~~~~
\frac{dC_m}{C_m^2}\,=\,
-\theta(t-t_1)\,
\beta\,a^{r-4}(t)\,dt\,=\,
-\theta\left[a(t)-a(t_1)\right]\,
\frac{\beta\,da}{a^{5-r}H}\label{ca3}
\end{eqnarray}
Eq.(\ref{ca3}) may be used to determine 
$C_m(t)$. This, in turn, may be used to find 
$\frac{d\tilde{\rho}_m(t)}{dt}$ from
(\ref{c1a1}), and $\rho_R$ from
(\ref{a21a}), and so
$\Delta\omega_R$ from
(\ref{a21d}). This will be done below
for a relatively simple, yet phenomenologically
relevant case.

Next we find $C_m(t)$ for some choices of $H$ by using
(\ref{ca3}).
For the sake of simplicity
consider the cases where
\begin{equation}
H\,=\,\xi\,a^s(t) \label{cn5}
\end{equation}
Here $\xi$ and $s$ are some constants. In fact
(\ref{cn5}) contains most of the phenomenologically 
relevant simple cases i.e. 
radiation, matter, cosmological constant
dominated eras. After using (\ref{cn5}),
(\ref{ca3}) becomes
\begin{eqnarray}
-\frac{1}{C_m(t)}
&=&
-\theta(t-t_1)\,
\frac{\beta}{\xi}\int\frac{\,da}{a^{5-r+s}}
-\frac{1}{C_m(t_1)} \nonumber \\
&&=\,
\frac{\beta\,\theta(t-t_1)}
{\xi(4-r+s)}\left[
\frac{1}{a^{4-r+s}(t)} \,-\,
\frac{1}{a^{4-r+s}(t_1)}\right] \,-\,
\frac{1}{\rho_{m0}}
\label{cn6}
\end{eqnarray}
where we have used $C_m(t_1)\,=\,\rho_{m0}$. 
(\ref{cn6}) results in
\begin{eqnarray}
&&C_m
\,=\,
\frac{\xi\,\rho_{m0}(4-r+s)
a^{4-r+s}(t_1)
a^{4-r+s}(t)}
{-\beta\rho_{m0}\theta(t-t_1)\,\left[
a^{4-r+s}(t_1)\,-\,
a^{4-r+s}(t)\right]
\,+\,\xi\,(4-r+s)
a^{4-r+s}(t_1)a^{4-r+s}(t)} \nonumber \\
&&\,=\,
\frac{\xi\,\rho_{m0}(4-r+s)}
{\xi\,(4-r+s)
\,+\,
\beta\rho_{m0}\theta(t-t_1)\,
\left[a^{r-s-4}(t_1)\,-\,
a^{r-s-4}(t)\right]}
\label{cn7}
\end{eqnarray}
Eq.(\ref{cn7}) implies that 
$s\,\rightarrow\,r-4$ limit is only well-defined for 
$\beta=0$ and is constant, and becomes 
not-well defined if $\beta\neq\,0$.
This suggests that at a universe where $s=r-4$ the processes
in Figure \ref{fig:1} can not take place. 
Moreover $C_m\,\rightarrow\,0$ as 
$a(t)\,\rightarrow\,\infty$ for $r-s-4\,>\,0$ while
$C_m$ goes a non-zero constant value  as 
$a(t)\,\rightarrow\,\infty$ for $r-s-4\,<\,0$. This 
implies that this process is not feasible in the case
$r-s-4\,<\,0$ for an expanding universe.

Eq.(\ref{cn7}) results in
\begin{eqnarray}
-\frac{d\tilde{\rho}_m(t)}{dt}
\,=\,\frac{\dot{C}_m(t)}{a^3(t)}
\,=\,
-\frac{\theta(t-t_1)\,\rho_{m0}^2\beta
\xi^2\,(4-r+s)^2a^{r-7}(t)}
{\{\xi\,(4-r+s)
\,+\,
\beta\,\rho_{m0}\,\theta(t-t_1)\,
\left[a^{r-s-4}(t_1)\,-\,
a^{r-s-4}(t)\right]\}^2}
\label{cn12}
\end{eqnarray}
\begin{eqnarray}
\rho_m(t)\,=\,\frac{C_m}{a^3}
\,=\,
\frac{\xi\,\rho_{m0}(4-r+s)\,a^{-3}(t)}
{\xi\,(4-r+s)
\,+\,
\beta\rho_{m0}\theta(t-t_1)\,
\left[a^{r-s-4}(t_1)\,-\,
a^{r-s-4}(t)\right]}
\label{cn13}
\end{eqnarray}

\subsubsection{Evolution of the energy densities through an example}

Now we consider an example to check if one may realize
the correct evolution of the total energy density in a consistent way. 
To check the possibility of the resonance obtained during conversion for 
matter to radiation that serve as 
dark energy we take the simplest choice i.e. the case where the resonance
particles induce a condensate that behaves as cosmological constant.
We consider, 
first, a matter dominated 
universe for 
$t_1\,<\,t\,<\,t_2$, 
where $s=-\frac{3}{2}$ (that may be taken as the universe just after the conversion of matter
particles to the resonance state),  
then a cosmological constant-like 
resonance dominated universe where $s=0$ for $t_2\,<\,t\,<\,t_3$, 
finally, the radiation dominated era at the end where $s=-2$ for $t_3\,<\,t$. For simplicity
we assume abrupt changes in the equations of states as one passes from one era to the other although in a
more realistic case the changes would be smooth. However this choice is enough to give the essentials points.
We take $r=7$ i.e. we take the intermediate state resonance be an extremely narrow resonance (to have a sufficiently long
lifetime).
In other words we consider the case
\begin{eqnarray}
&&\omega_R\,=\,-1~~~,~~r\,=\,7~~~~~\mbox{and}\nonumber \\
&&\mbox{Initially}~~s\,=\,-\frac{3}{2}
~,~~~\mbox{in between}~~
s\,=\,0~,~~~\mbox{at late times}~~s\,=\,-2
\label{cn14}
\end{eqnarray}
We will assume that $\beta$ is constant in all eras. 
We give $C_m$, $\rho_R$, $C_r$ in these eras below by using the formulas 
derived above. 
(Please refer to Appendix B for the details of the derivations).
$\rho_m$ and $\rho_r$ are evident from
$C_m$ and $C_r$. 

$C_m(t)$ at different eras are found as
\begin{eqnarray}
&&
\mbox{For}~~t_1\,<\,t\,<\,t_2~~~~~~~
C_m
\,=\,
\frac{\rho_{m0}}
{1\,+\,
\frac{2\beta\rho_{m0}}{9\xi_1}\,
\left(a^\frac{9}{2}\,-\,
a_1^\frac{9}{2}\right)}
\label{cn7a} 
\end{eqnarray}
where $a=a(t)$, $a_1=a(t_1)$, and
(\ref{cn7a})
is directly found from (\ref{cn7})
for $r=7$, $s=s_1=-\frac{3}{2}$, 
$\xi=\xi_1$. 
\begin{eqnarray}
&&\mbox{For}~~t_2\,<\,t\,<\,t_3\nonumber\\
&&~~C_m(t)\,=\,\rho_{m0}\{
1\,+\,
\rho_{m0}\beta[
\frac{2}{9}\xi_1^{-1}
(a_2^\frac{9}{2}\,-\,
a_1^\frac{9}{2})
\,+\,\frac{1}{3}\xi_2^{-1}
(a^3 \,-\,
a_2^3)]\}^{-1}
\label{cn19b}
\end{eqnarray}
where $a_2=a(t_2)$, and
to find $C_m$ in this case we have divided
the integral in (\ref{cn6}) into two parts; one for the era when
$s=-\frac{3}{2}$ between $t_1$ and $t_2$, and then the era when
$s=0$, 
$\xi=\xi_2$. 
between $t_2$ and $t$. In a similar way as in (\ref{cn19b})
we find
\begin{equation}
\mbox{For}~~t_3\,<\,t~~~~~~~~
C_m(t)\,=\,\frac{\rho_{m0}}{A^\prime\,+\,
\,B^\prime\,a^5}
\label{cn21b}
\end{equation}
Here $A^\prime$, $B^\prime$ are some constant whose explicit forms
are given in (\ref{cn21bbA}).
We notice that $C_m$ starts from $\rho_{m0}$ and goes to 
zero as $a\,\rightarrow\,\infty$ as expected.

$\rho_R$'s in these eras are found as
\begin{eqnarray}
&&\mbox{For}~~t_1\,<\,t\,<\,t_2 ~~~~~~~
\rho_R(t)\,=\,
\,e^{-\Gamma(t-t_1)}\frac{C_R}{a^3}
\,\simeq\,
\frac{2}{3}\,
\frac{\rho_{m0}^2\beta}
{\xi_1}\frac{1}{a^3(t)}
\left[a^{\frac{3}{2}}\,-\,a_1^\frac{3}{2}\right]
\label{cn19} \\
&&\mbox{For}~~t_2\,<\,t\,<\,t_3\nonumber \\
&&~~\rho_R(t)\,=\,
\left(\frac{a_1}{a}\right)^3\frac{1}{3}
\,\left[\frac{2\rho_{m0}^2\beta}
{\xi_1}\{a_2^{\frac{3}{2}}\,-\,
a_1^\frac{3}{2}\}
\,+\,
\frac{\rho_{m0}^2\beta}{B\xi_2}\{\frac{1}{A}\,-\,
\frac{1}{A\,+\,
B\,a^3}\}\right]
\label{cn20d} 
\end{eqnarray}
where have taken
$\gamma\,=\,\frac{\Gamma}{H}\,=\,3$ in
(\ref{cn20bA}) as a generic case where the 
integration is simpler, and $A$, $B$ are some
constants whose explicit forms are given in (\ref{cn20aaA}) and
(\ref{cn20abA}).
For $t\,>\,t_3$
\begin{eqnarray}
&&\rho_R(t)
\,=\,e^{-\Gamma(t-t_1)}\,C_R(t) \nonumber \\
&&\simeq\,
\frac{4096\xi_3^4\rho_{m0}^2\beta}{3e^4\,\Gamma^4\xi_1}
\left(a^2-a_1^2\right)^{-4} 
\left[2\left(a_2^\frac{3}{2}\,-\,
a_1^\frac{3}{2}\right)\,
\,+\,
\frac{a_1^\frac{3}{2}}{B}
\left(1\,-\,
\frac{1}{1\,+\,B\,a_3^3}\right)\right]
\label{cn28}
\end{eqnarray}
where we have essentially 
employed (\ref{a2aar}), (\ref{a21}),
(\ref{a21a}), and (\ref{cn7a}),
(\ref{cn19b}), (\ref{cn21b}).
One notices that $\rho_R$ starts from zero at
$t=0$ and keeps rising at intermediate times 
and goes to zero due to its decay as 
$a\,\rightarrow\,\infty$ as expected. In fact one
may also refer to the behavior of $C_R$ in Appendix B
to see that $C_R$ is zero first and goes to a
constant value as
$a\,\rightarrow\,\infty$. $\rho_R$ would behave like
a cosmological constant at large scales factors if it had not decayed.

The corresponding $C_r$'s are
\begin{eqnarray}
&&\mbox{For}~~t_1\,<\,t\,<\,t_2 \nonumber \\
C_r(t)&=&
\frac{2\Gamma\rho_{m0}^2\beta}{\xi_1^2}
\int_{a_1}^a\,(a^{\prime\,3}\,-\,
a_1^\frac{3}{2}a^{\prime\,\frac{3}{2}})\,da^\prime
\,=\, 
\frac{\Gamma\rho_{m0}^2\beta}{2\xi_1^2}\left[a^4\,-\,
\frac{8a_1^\frac{3}{2}}{5}a^\frac{5}{2}\,+\,\frac{3}{5}a_1^4\right]
\label{cn19cr}
\end{eqnarray}
\begin{eqnarray}
&&\mbox{For}~~t_2\,<\,t\,<\,t_3 \nonumber \\
C_r(t)&\simeq&
\frac{\Gamma\rho_{m0}}{\xi_1}
\left[\frac{\rho_{m0}\beta}{2\xi_1}
\left(a_2^4-
\frac{8a_1^\frac{3}{2}}{5}
a_2^\frac{5}{2}
+\frac{3}{5}a_1^4\right)
\,+\,
\left(\frac{2\rho_{m0}\beta\,a_1^3(a_2^\frac{3}{2}-
a_1^\frac{3}{2})
}{3\xi_2}
\right)
(a-a_2)\right]
\label{cn21cr}
\end{eqnarray}
\begin{eqnarray}
&&
\mbox{For}~~t_3\,<\,t \nonumber \\
C_r(t)&\simeq&
\frac{\Gamma\rho_{m0}}{\xi_1}
\left[\frac{\rho_{m0}\beta}{2\xi_1}
\left(a_2^4-
\frac{8a_1^\frac{3}{2}}{5}
a_2^\frac{5}{2}
+\frac{3}{5}a_1^4\right)
\,+\,
\left(\frac{2\rho_{m0}\beta\,a_1^3(a_2^\frac{3}{2}-
a_1^\frac{3}{2})
}{3\xi_2}
\,+\,
a_1^\frac{9}{2}\right)
(a_3-a_2)\right] \nonumber \\
&&+\,C^{\prime\prime}\,
\left[
\frac{a_1^2\,-\,3a_1^2a^2\,+\,3a^4}{6(a^2-a_1^2)}\,-\,
\frac{a_1^2\,-\,3a_1^2a_3^2\,+\,3a_3^4}{6(a_3^2-a_1^2)}
\right]
\label{cn28cr}
\end{eqnarray}
where $C^{\prime\prime}$ is a constant whose explicit 
form is given in (\ref{cn28craA}), 
and $C_r$'s are found by using 
(\ref{a2cr1}) and (\ref{cn19}),
(\ref{cn20d}), (\ref{cn28}).
We notice that $C_r$ starts from zero and
goes to a constant value 
as $a$ increases as expected.

\subsubsection{A concrete particle physics model}

In this sub-subsection, 
through a concrete particle physics model,
we illustrate the main lines of the results of our analysis on
the use of on-shell intermediate state particle condensates 
for cosmic accelerated expansion. In fact many scalar fields
in cosmology including
inflaton or quintessence fields may be considered as 
Bose-Einstein condensates
in the light of their homogeneity
and coherence \cite{Dolgov}. This approach is elaborated and detailed in many
models of dark matter \cite{Urena,DM-BEC}, and a few models of dark energy 
\cite{Morikawa-1,Morikawa-2,DE-BEC-other} and inflation 
\cite{inflation-BEC1,inflation-BEC2}. We study a specific model of dark energy 
similar to a model in literature \cite{Morikawa-2} 
to see how the formulation developed
in the previous parts of this subsection provides additional insight and improvement. 
The 
model we consider is simple yet sophisticated enough to be able to study main 
lines of this scheme in a proper way.   

We consider the following Lagrangian 
\begin{eqnarray}
{\cal 
L}&=&\sqrt{-g}\{
-g^{\mu\nu}\left(\partial_\mu\phi^*\partial_\nu\phi\,+\,
\partial_\mu\chi_{m1}^*\partial_\nu\chi_{m1}\,+\,
\frac{1}{2}\partial_\mu\chi_{m2}\partial_\nu\chi_{m2}\,+\,
\partial_\mu\chi_{r1}^*\partial_\nu\chi_{r1}\,+\,
\frac{1}{2}\partial_\mu\chi_{r2}\partial_\nu\chi_{r2}\right)
\nonumber \\
&&-\,m_\phi^2\phi^*\phi\,-\,\frac{\lambda}{2}(\phi^*\phi)^2
\,-\,m_{m1}^2\chi_{m1}^*\chi_{m1}
\,-\,\frac{1}{2}m_{m2}^2\chi_{m2}^2
\,-\,m_{r1}^2\chi_{r1}^*\chi_{r1}
\,-\,\frac{1}{2}m_{r2}^2\chi_{r2}^2 \nonumber \\
&&-\,\mu_m\phi^*\chi_{m1}\chi_{m2}
\,-\,\mu_m^*\chi_{m1}^*\phi\chi_{m2}
\,-\,\mu_r\phi^*\chi_{r1}\chi_{r2}
\,-\,\mu_r^*\chi_{r1}^*\phi\chi_{r2}\}
\label{pp1}
\end{eqnarray}
Here $\phi$, $\chi_{m1}$, $\chi_{r1}$ are complex scalars while
$\chi_{m2}$, $\chi_{r2}$ are real scalars. These are the particles
in the two body scattering process in Figure \ref{fig:1}, namely,
$\chi_{m1}$, $\chi_{m2}$ denote the incoming matter particles with momenta 
$p_1$, $p_2$
while 
$\chi_{r1}$, $\chi_{r2}$ denote the outgoing radiation
particles with momenta $p_3$, $p_4$, 
and $\phi$ denotes the on-shell intermediate state particle in the figure.

The Klein-Gordon equation for $\phi$ is obtained from (\ref{pp1}) as
\begin{eqnarray}
&&\frac{\partial^2\phi}{\partial\,t^2}\,+\,3H\frac{\partial\phi}{\partial\,t}
\,-\,\frac{c^2}{a^2}\Delta\,\phi\,+\,\frac{m_\phi^2c^4}{\hbar^2}\phi\,+\,c^2\lambda\phi(\phi^*\phi)
\,+\,c^2\mu_m\chi_{m1}\chi_{m2}\,+\,c^2\mu_r\chi_{r1}\chi_{r2}\,=\,0 \nonumber \\
\label{pp2}
\end{eqnarray}
where $\Delta$ is the 3-dimensional Laplace operator, we have assumed the metric to be the spatially flat case of 
(\ref{b1}) i.e. we have 
assumed the perturbations of $\phi$ due to its motion be negligible with respect to 
the background, and we have introduced $c$ and $\hbar$ into the equation that has 
been taken as $c=\hbar=1$ in (\ref{pp1}) to see the reduction of (\ref{pp2}) 
to Schr{\"{o}}dinger equation in the non-relativistic limit more clearly. 

To obtain the non-relativistic limit of (\ref{pp2}) (to compare this case 
with the standard literature at atomic scales 
where cosmic expansion is not considered) we assume that the second 
term is negligible with respect to the others. 
One may neglect the second term 
provided that $\frac{m_\phi\,c^2}{\hbar}\,\gg\,3H$ (that may be seen more easily after transforming
$\phi$ to $\psi$ as given below). 
Consider the transformation 
$\phi\,=\,e^{-i\frac{m_\phi\,c^2}{\hbar}t}\psi$. 
Note that this transformation essentially subtracts the rest mass energy from the total energy of $\phi$.  
If $\phi$ is non-relativistic for a long time then its time evolution should be slow. 
These together, in turn, imply $|\frac{\ddot{\psi}}{\dot{\psi}}|\,\ll\,\frac{2M_\phi\,c^2}{\hbar}$. 
Then (\ref{pp2}) reduces to
\begin{equation}
-\,i\hbar\dot{\psi}\,-\,\frac{\hbar^2}{2a^2m_\phi}\Delta\,\psi\,+\,\tilde{\lambda}\psi(\psi^*\psi)
\,+\,\tilde{\mu}_m\tilde{\chi}_{m1}\chi_{m2}\,+\,\tilde{\mu}_r\tilde{\chi}_{r1}\chi_{r2}\,=\,0 \label{pp4}
\end{equation}
Here
$\psi=e^{i\frac{m_\phi\,c^2}{\hbar}t}\phi$, 
$\tilde{\chi}_{m1}=e^{i\frac{m_\phi\,c^2}{\hbar}t}\chi_{m1}$, 
$\tilde{\chi}_{r1}=e^{i\frac{m_\phi\,c^2}{\hbar}t}\chi_{r1}$, $\tilde{\lambda}=\frac{\hbar^2}{2m_\phi}\lambda$, 
$\tilde{\mu}_m=\frac{\hbar^2}{2m_\phi}\mu_m$, $\tilde{\mu}_r=\frac{\hbar^2}{2m_\phi}\mu_r$. 
Eq.(\ref{pp4}) is the Gross-Pitaevskii equation \cite{Pethick,Pitaevskii} for the metric in (\ref{b1})
where the trapping 
potential is zero and there is an external source term \cite{external-source}, namely, 
$\mu_m\tilde{\chi}_{m1}\chi_{m2}\,+\,\mu_r\tilde{\chi}_{r1}\chi_{r2}$. 
Note that Gross-Pitaevskii equation 
is the equation for a particle in a condensate, so it is the 
equation for the condensate state since all particles are in the same state in a Bose-Einstein condensate. 
The external source 
term in (\ref{pp4}) accounts for the conversion of $\chi_{m1}$, $\chi_{m2}$ to $\psi$ and the decay of $\psi$ to $\chi_{r1}$ 
and $\chi_{r2}$. 

We let the coupling constants $\mu_m$ and $\mu_r$ satisfy 
$\mu_m\,\gg\,\mu_r$ since we want a quick formation and a long lifetime for the condensate to simplify the situation. 
A short condensate formation time together with the condition that 
$\chi_{m1}$, $\chi_{m2}$ are non-relativistic and the mass of $\phi$ being comparable to those of  
$\chi_{m1}$, $\chi_{m2}$ guarantees smallness of any entropy production through the condensation process 
\cite{Dolgov,non-adiabatic-BEC}. To be specific; the short duration of the 
condensation and small velocities guarantee smallness of any work done during the process, and the non-relativistic nature of 
$\chi_{m1}$, $\chi_{m2}$, $\phi$ 
guarantees the smallness of any heat production, so any probable entropy production due to mismatch of these terms will be small. 
In fact in the case of Bose-Einstein condensation in atomic gases the formation time of the 
condensate is very small compared to its lifetime
\cite{formation-time,full-BEC,BEC-lifetime}. 
Note that the 
condensates described by Gross-Pitaevskii equation for $\lambda\,>\,0$ correspond to an ideal case at $T\,=\,0$ where there are
no inelastic two and three body collisions in the condensate that limits its life-time \cite{BEC-lifetime}, hence a condensate 
described 
by Gross-Pitaevskii equation form more easily and is stable i.e. has an infinite lifetime. Therefore the assumption of a short time 
for the formation of the condensate in this case is a reasonable assumption.  
Moreover we require a long lifetime to obtain a simple model compatible with the present energy densities of dark energy and dark 
matter as we will see at the end of this sub-subsection. This requirement is easily satisfied provided that we take $\mu_r$ 
sufficiently small.
The equations 
(\ref{pp1}) and (\ref{pp4}) are similar to those in \cite{Morikawa-2} 
except the additional scalars $\chi_{mi}$, $\chi_{ri}$, i=1,2; 
hence the presence of an external source term in this case. 
We let 
$m_{m1}^2\chi_{m1}^*\chi_{m1}\,\gg\,\mu_m\chi_{m1}^*\phi\chi_{m2}$,
$m_{m2}^2\chi_{m2}^*\chi_{m2}\,\gg\,\mu_m\chi_{m1}^*\phi\chi_{m2}$ so that one may consider
$\chi_{m1}$, $\chi_{m2}$ as (almost) free particles, and in their non-relativistic limits
their equations of state become (almost) zero, so considering them as matter particles
is insured. In a similar way we let
$m_{r1}^2\chi_{r1}^*\chi_{r1}\,\gg\,\mu_m\chi_{r1}^*\phi\chi_{r2}$,
$m_{r2}^2\chi_{r2}^*\chi_{r2}\,\gg\,\mu_m\chi_{r1}^*\phi\chi_{r2}$
so that one may consider
$\chi_{r1}$, $\chi_{r2}$ as (almost) free particles, and hence we may consider them as radiation 
provided that and $m_{r1}^2\simeq\,0$, $m_{r2}^2\simeq\,0$.
Another important difference 
between \cite{Morikawa-2} and this one is that 
\cite{Morikawa-2} takes $\lambda\,<\,0$ while we 
take $\lambda\,>\,0$ which is the case in the original Gross-Pitaevskii equation. This makes the model simpler since the condensate 
in this case becomes stable (in the absence of the $\phi^*\chi_{1r}\chi_{2r}$ term) while in \cite{Morikawa-2} the condensate 
experiences few collapse and re-formation cycles. In fact the condensate in this case (even in the presence of the 
$\phi^*\chi_{1r}\chi_{2r}$ term) is rather stable since we take $\mu_r$ very small.
Another potential point that makes the condensate more stable is the use of complex scalars i.e. charged scalars that makes the 
condensate more stable than a naive expectation \cite{Dolgov}.
We assume that the particle number density is sufficiently high enough so that the de Broglie wave length 
$\lambda_{dB}\,=\,\sqrt{\frac{2\pi\hbar^2}{mkT}}$ is larger than the mean separation $n^{-\frac{1}{3}}$ between the particles. 
This condition may be converted to an upper bound on the temperature of $\phi$'s 
\begin{equation}
T\,<\,T_c\,=\,\frac{2\pi\hbar^2n^{\frac{2}{3}}}{k\left(\zeta(\frac{3}{2})\right)^\frac{2}{3}m_\phi} \label{pp5}
\end{equation}
where $k$ is the Boltzmann constant, $n$ is the number density. In the case where the $\phi$'s were 
in thermal equilibrium with baryonic matter
and radiation at some initial time, Eq.(\ref{pp5}) puts an upper bound on $m_\phi$ while there is no bound on $m_\phi$ in the case 
where $\phi$'s never had a thermal 
equilibrium with baryonic matter and radiation. 
In the case of non-relativistic $\phi$'s the temperature of $\phi$'s remain below $T_c$ at later 
times once it is below $T_c$ at some initial time in the past \cite{Morikawa-2}. In this study we assume that $\phi$'s are 
non-relativistic and they had never thermal equilibrium with baryonic matter and radiation in the past. 

After the basic elements of condensate formation for (\ref{pp1}), 
now we are ready to derive its
results at macroscopic level by using the general analysis we have given. In another words we will relate
the microscopic view (i.e. the particle physics view) of the condensate given above 
to its macroscopic view at the level of energy densities in the light of the analysis 
we have given before this sub-subsection. 
The evolution of the energy densities may be derived from (\ref{a2aar}), (\ref{a21c}), 
and (\ref{c1a1}), (\ref{r1b}), as
\begin{eqnarray}
&&H^2\,=\,\frac{8\pi\,G}{3}\left(\rho_m\,+\,\rho_\phi\,+\,\rho_r\right) \label{pp5aa} \\
&&\dot{\rho}_m\,+\,3H\rho_m\,=\,-\frac{d\tilde{\rho}_m}{dt}\,=\,-\beta\,\left(a(t)\right)^{r-1}\rho_m^2 \label{pp5a} \\  
&&\dot{\rho}_\phi\,+\,3H(1+\omega_\phi)\rho_\phi\,=\,
\frac{d\tilde{\rho}_m}{dt}\,-\,
\Gamma\rho_\phi \,=\,  
\beta\,\left(a(t)\right)^{r-1}\rho_m^2 
\,-\,\Gamma\rho_\phi \label{pp5b} \\  
&&\dot{\rho}_r\,+\,4H\rho_m\,=\,
\frac{d\tilde{\rho}_r}{dt}\,=\,
\Gamma\rho_\phi \label{pp5c} 
\end{eqnarray}  
where 
$\rho_m$ consists of $\chi_{m1}$ and $\chi_{m2}$, $\rho_r$ consists of
$\chi_{r1}$ and $\chi_{r2}$, and $\rho_\phi$ consists of the $\phi$ particles. 
Here we do not make any distinction between the resonance particles
produced as the intermediate state of the 4-point function in Fig. \ref{fig:1} and the
particles in the condensate because we assume that the condensate is formed almost simultaneously
after the formation of the resonance particles as we have mentioned before. 
The equations (\ref{pp5a}), (\ref{pp5b}) somewhat differ from those of \cite{Morikawa-2} since the   
$\frac{d\tilde{\rho}_m}{dt}$ term in the above equations is proportional to $\rho_m$
in \cite{Morikawa-2} while it is proportional to $\rho_m^2$ here. 
Taking 
$\frac{d\tilde{\rho}_m}{dt}$ proportional to $\rho_m^2$ (due to (\ref{c1a1}) 
seems more correct in the view
that Bose-Einstein condensation involves interaction of particles in a Bose gas rather 
than their decay (where
decay rate would be proportional to the energy density of the decaying 
particles). \cite{Morikawa-2} also considers a possible 
dependence of the decay widths on the 
scale factor in an ad hoc way. While the dependence of
$\frac{d\tilde{\rho}_m}{dt}$ on scale factor in this study and 
in \cite{Morikawa-2} are similar their origins are different.
The origin of the dependence of
$\frac{d\tilde{\rho}_m}{dt}$ in \cite{Morikawa-2} is through 
$\Gamma$'s while 
in our case it results from the propagator part 
of the cross section, not through $\Gamma$'s.
We have found that dependence of the self energies on the 
momentum in perturbative quantum field theory
suggests that the leading order contribution to $\Gamma$ is independent of scale factor. 
Moreover the explicit
form of $\rho_m$ in terms of the cross section is derived in this study. 

One may see the general lines of the formation of the condensate
by following a procedure similar to that of \cite{Morikawa-2}.
Eq.(\ref{pp5b}) be written as
\begin{equation}
\dot{\rho}_\phi\,+\,6H
\left[\dot{\phi}^*\dot{\phi}+
\frac{1}{a^2}
(\vec{\nabla}\phi^*).(\vec{\nabla}\phi)
\right]\,=\,
\beta\,\left[a(t)\right]^{r-1}\rho_m^2 
\,-\,\Gamma\rho_\phi \label{pp6} 
\end{equation}
Here we have used
$\rho_\phi+p_\phi=2\left[\dot{\phi}^*\dot{\phi}\,+\,
\frac{1}{a^2}
(\vec{\nabla}\phi^*).(\vec{\nabla}\phi)\right]$. Note that in a case where $\phi$ is 
homogenous we would have $\vec{\nabla}\phi=0$ while here we take it non-zero at initial times of the
formation of the condensate (although the background is taken as homogeneous). 
Otherwise there will be no condensate formation starting from $\rho_\phi\,=\,0$
at initial time $t_1$
since
\begin{equation}
\rho_\phi\,=\,
\dot{\phi}^*\dot{\phi}\,+\,
\frac{1}{a^2}
(\vec{\nabla}\phi^*).(\vec{\nabla}\phi)\,+\,V \label{pp6a} 
\end{equation}
implies
\begin{equation}
\phi(t_1)\,=\,0\,,~~\dot{\phi}(t_1)\,=\,0\,,~~\vec{\nabla}\phi(t_1)\,=\,0 
\label{pp7a}
\end{equation}
because 
\begin{equation}
V\,=\,
\,m_\phi^2\phi^*\phi\,+\,\frac{\lambda}{2}(\phi^*\phi)^2\,+\,
\mu_m\phi^*\chi_{m1}\chi_{m2}
\,+\,\mu_r\phi^*\chi_{r1}\chi_{r2}
\,+\,h.c.
\label{pp7}
\end{equation}
(where $h.c.$ implies taking the Hermitean conjugate of the proceeding terms) is positive
for all values of $\phi$ since we take all coefficients positive. (In fact even in the
case where $\mu_m$, $\mu_r$ are negative $V$ is still positive since we take the corresponding terms
very small compared to the rest of the terms). 
These together with 
Eq.(\ref{pp2}) imply that we 
should have 
\begin{equation}
\Delta\phi(t_1)\,>\,0~~\mbox{and}~~~
\frac{\partial^2\phi(t_1)}{\partial\,t^2}\,>\,0 \label{pp7b}
\end{equation}
During the formation of the condensate one must
have $\dot{\rho}_\phi\,>\,0$. Therefore Eq. (\ref{pp6}) implies that 
during the formation of the condensate one has one must have
\begin{equation}
\beta\,\left(a(t)\right)^{r-1}\rho_m^2 
\,-\,\Gamma\rho_\phi\,>\, 6H(\rho_\phi-V)\,=\,6H\left(|\dot{\phi}|^2\,+\,
\frac{1}{a^2}
|\vec{\nabla}\phi|^2\right)\,>\,0 \label{pp8} 
\end{equation}
$\dot{\phi}$, $\vec{\nabla}\phi$ start with the values given in (\ref{pp7a}) 
and increase because of (\ref{pp7b}) till the bound in (\ref{pp8}) is saturated.
Eq.(\ref{pp2}) implies that first 
$\frac{\partial^2\phi}{\partial\,t^2}$ 
becomes zero i.e. $\dot{\phi}$ reaches its
maximum value while 
$\vec{\nabla}\phi$ keeps rising i.e. still  
$\Delta\phi\,>\,0$. 
$\vec{\nabla}\phi$ keeps rising till the bound in 
(\ref{pp8}) is saturated once again. (Note that, as the right hand side of the
inequality in (\ref{pp8}) increases its left hand side decreases by time since the
energy density of $\rho_m$ gets smaller, that of $\rho_\phi$ gets larger.)
Then 
$\Delta\phi$ changes sign and 
$|\vec{\nabla}\phi|^2$ decreases too till it eventually becomes zero, so the right hand-side (\ref{pp8}) becomes zero.
Finally the left-hand side of (\ref{pp8}) becomes zero too by evolution of the energy densities, and the final values of
$\rho_m$ and $\rho_\phi$ are reached (provided that the decay of $\phi$ is neglected i.e. the term with coefficient $\mu_r$ is 
neglected since it very small). Therefore 
one may get a rough picture of the 
evolution of the energy densities. 

Now let us see if this model is compatible with the present day energy densities, 
at least, at an order of magnitude level. We will identify the condensate by dark energy
and the leftover part of $\rho_m$ after formation of the condensate 
by dark matter. We will neglect the contributions due to baryonic matter and (the observable) radiation for simplicity. 
We take the time that passed since the start of 
the condensation $t_1$ till the present time $t_0$ be much smaller than the lifetime of the condensate i.e. we let 
$(t_0-t_1)\,\ll\,\Gamma^{-1}$ to ensure that the energy density of $\chi_{r1}$ and $\chi_{r2}$ is below the observational 
bounds on dark radiation \cite{dark-radiation}. This condition is expressed in a more concrete form in the next equation. To derive 
that equation we make use of the following approximations and equations: 
We take $\rho_\phi$ almost constant 
from $t_1$ to $t_0$, and make use of (\ref{a1b}) and
(\ref{a2cr1}) (where $\rho_R=\rho_\phi$). 
We approximate the integral in (\ref{a2cr1}) as the sum of two parts; the first part being from $t_1$ to $t_q$ where we 
assume the universe was wholly matter dominated and from $t_q$ to $t_0$ where we assume that the universe was wholly 
$\rho_\phi$ dominated which is a rough approximation to the evolution of the universe. We neglect the 
radiation dominated era because we neglect the contribution of (the observable) radiation in this study. (In fact even we had 
included 
the contribution of the usual observable radiation its effect would turn out to be negligible in the integral given below.) 
Then (\ref{a2cr1}) and (\ref{a1b})
imply 
\begin{eqnarray} 
&&
\frac{\frac{\Omega_\gamma}{10}}{\Omega_{DE}}
\,\simeq\,10^{-5} \,>\,
\frac{\Omega_{r}}{\Omega_{DE}}
\,=\,\frac{\rho_r(t_0)}{\rho_\phi(t_0)}\,\simeq\,\Gamma\,
\int_{t_1}^{t_0}
a^4(t^\prime)
\theta(t-t^\prime)\,dt^\prime \nonumber \\
&&\simeq\,\frac{\Gamma}{H_0}\left[
\int_{a(t_1)}^{a(t_q)}
\frac{\left(a^\prime\right)^\frac{9}{2}}{\sqrt{\Omega_{DM}}}da^\prime\,+\,
\int_{a(t_q)}^{a(t_0)}
\frac{\left(a^\prime\right)^3}{\sqrt{\Omega_{DE}}}da^\prime\right]
\,\simeq\,0.3\frac{\Gamma}{H_0}
\,\simeq\,0.3\,t_0\,\Gamma
\label{pp01}
\end{eqnarray}
where 
we used the fact that 
the present energy density of a radiation component other than photons and neutrinos, $\rho_{r}$
is, at most, at the level of a tenth of the
energy density of photons $\rho_\gamma$ \cite{dark-radiation};
$\rho_{DE}$ is the present energy density for dark energy; 
$\Omega$'s denote the corresponding density 
parameters, and we have used the numerical values of the density parameters in \cite{PDG} where $\Omega_{DE}=\Omega_\Lambda$ and that 
the time for the beginning of the dark energy dominated era is at about the redshift $z=0.65$ i.e. at $a(t_q)$=0.6.
One may obtain a similar result by using a more heuristic approach as well:
Provided that $\Gamma\,(t_0-t_1)\,\ll\,1$,
the ratio of the energy 
density of the decayed part of the 
condensate to its energy density at present 
$\frac{\rho_{r0}}{\rho_{DE0}}$
roughly satisfies
\begin{equation}
\frac{\frac{\Omega_\gamma}{10}}{\Omega_{DE}}
\,\simeq\,10^{-5} \,>\,
\frac{\rho_{r0}}{\rho_{DE0}}\,=\,
\frac{1-e^{-\Gamma\,(t_0-t_1)}}{e^{-\Gamma\,(t_0-t_1)}}\,\simeq\,
\Gamma\,(t_0-t_1)
\label{pp9}
\end{equation}
In the notation of the 
preceding sub-subsection we have $t_2\,<\,t_0\,\ll\,t_3$. If  $\phi$ never had  
a thermal equilibrium with baryonic matter
and photons 
then the condensation must have started before
matter dominated era if we want to mimic $\Lambda$CDM (because of its observational success). 
On the other hand, if $\phi$ had thermal equilibrium with baryonic matter and photons in 
the past then $\phi$ must have decoupled from
baryonic matter and radiation before the time of nucleosynthesis to keep the successful 
nucleosynthesis scenario of the standard model intact. 
In either case $t_0-t_1$ must be in the order of the age of the universe i.e. in the order of $10^{17}$ seconds. 
Hence (\ref{pp01}) or (\ref{pp9}) 
implies that $\frac{1}{\Gamma}\,>\,10^{22}$ seconds. In the light of these observations we derive a rough lower bound on the value of
the production cross section of $\phi$'s by using (\ref{pp8}). Since $\frac{1}{\Gamma}\,>\,10^{22}$ seconds, and 
after the end of the formation process of the condensate 
(as discussed after (\ref{pp8})) 
we have $\beta\left(a(t)\right)^{r-1}\rho_m^2\,-\,\Gamma\rho_\phi\,=\,0$,  
we obtain
\begin{equation}
\frac{1}{10^{22} sec.\,\times\,\beta}\,<\,
\frac{\Gamma}{\beta}
\,=\,\frac{\rho_m(t_0)}{\rho_\phi(t_0)}\rho_m(t_0)
\simeq\,
\left(\frac{0.26}{0.69}\right)\times\,0.26\,\rho_{crit}\,\simeq\,0.1\,\rho_{crit}
\label{pp10}
\end{equation}
Here $\rho_{crit}$ is the critical energy density at present. After using (\ref{pp10}) and the expression for $\beta$ in (\ref{c1a1}) 
and
the numerical value of $\rho_{crit}$
we may write
\begin{equation}
\beta\,=\,
\frac{\beta^\prime\,\sigma(t_1)\,v(t_1)}{E\,a(t_1)}\,>\,
(\frac{16}{9})\times\,10^{-25}\,(eV)^{-1}(sec)^{-1}
\label{pp11}
\end{equation}
i.e.
\begin{equation}
\frac{\sigma(t_1)}{cm^2}\,>\,\left(\frac{E}{eV}\right)\left(\frac{cm/sec}{v(t_1)}\right)\,\frac{a(t_1)}{\beta^\prime}
\left(\frac{16}{9}\times\,10^{-25}\right)
\label{pp12}
\end{equation}
It is evident from (\ref{pp12}) that one can produce the values of
the present dark matter  and dark energy density parameters 
$\Omega_{DM}\,\simeq\,0.26$ and 
$\Omega_{DE}\,\simeq\,0.69$ 
for a wide range of cross sections. For 
example if we take
$v(t_1)\,=\,\frac{1}{3}\times\,10^{-4}\,c$,
$a(t_1)\,=\,10^{-10}$, 
$\beta^\prime\,=\,\frac{1}{10}$, $\left(\frac{E}{eV}\right)\,=\,1$ then we find 
$\sigma(t_1)\,>\,\frac{16}{9}\,\times\,10^{-42}\,cm^2$, and $\sigma(t_1)\,>\,10^{-26}\,cm^2$ if  
$\left(\frac{E}{eV}\right)\,=\,10^9$,  
$a(t_1)\,=\,10^{-3}$ and all other parameters kept the same. 
We may also check if the formula we have derived for $\rho_R$ in this section 
is consistent with a constant value for 
$\rho_\phi$ at present 
provided that we adopt the values of 
parameters given above. In this case one can not directly use the results of the preceding sub-subsection because in this case 
$\gamma\,=\,\frac{\Gamma}{H}\,\ll\,1$ 
while in (\ref{cn20d}) (or in (\ref{cn20dA})) we had taken $\gamma=3$ to simplify the integrals 
for $\rho_r$. Instead we should go back to the original equation in (\ref{cn20aA}) 
and make use of the fact that 
$\frac{\rho_{m0}\beta}{\xi_i}\,\ll\,1$, $i=1,2$ in this case since $\xi_i$ are in the order of $H_0$. Therefore 
$\left(\frac{d\tilde{\rho}_m(t)}{dt}\right)_2\,\simeq\,\theta(t-t_2)\,\rho_{m0}^2\beta$. Then we insert this value in 
(\ref{cn20bA}) 
while we keep the exponential factor, and take the integral, and finally we use $\gamma\,=\,\frac{\Gamma}{H}\,\ll\,1$. Hence 
in this case $C_R$ is essentially constant at present, so the same is true for $\rho_R\,=\,\rho_\phi$.

\subsection{Some additional remarks for an intermediate 
state resonance}

In the preceding subsection we have given a particle physics description of 
an intermediate state with negative equation of state 
in the conversion of matter to radiation 
in terms of resonance particles that form a Bose-Einstein
condensate. We have also considered a concrete model in this context to
see the situation more explicitly. However such a specific model is not enough
to see all aspects of these types of models. We have neglected the era
for formation of the condensate since it is expected to be very small
compared to its lifetime and to be able focus on the more relevant points. 
A similar situation is true for the era of the decay of the condensate 
since this era corresponds to far future in our model, so phenomenologically
is less interesting. Although the study of these eras in detail is 
not essential in the context of the model we have considered for late time cosmic
acceleration it would be
relevant in models of inflation, especially in those where the condensate decays into
usual radiation rather than to dark radiation. Therefore in the following paragraphs
we give the overall picture
in terms of some simple generic hypothetical energy distribution to see the
overall picture more clearly.

After adding (\ref{a2aar}),
(\ref{a2bar}), (\ref{a21b}), and using
(\ref{r1a}) one obtains
\begin{eqnarray}
&&\dot{\rho}(t)\,+\,
3H\left[1\,+\,
\frac{1}{3}\frac{\rho_r(t)}{\rho(t)}\,+\,
\omega_R\frac{\rho_R(t)}{\rho(t)}
\right]\rho(t)\,=\,0
\label{ra3a} \\
&&\rho(t)\,=\,\rho_m(t)\,+\,\rho_r(t)\,+\,
\rho_R(t)\nonumber
\end{eqnarray}
Now we consider some specific cases in the light of the 
equation (\ref{ra3a}). 
Note that 
$\frac{\rho_R}{\rho}$ may be close to 1, especially if 
we identify $\rho_R$ by a resonance or a condensate 
formed as an intermediate state in the transition of 
matter to radiation as in the second graph in Figure 
\ref{fig:1}. Therefore the effect of $\rho_R$ 
may be sizeable for some time interval in the evolution 
of the 
universe. Moreover, since $\rho_R$ corresponds to 
some intermediate state in the conversion of matter 
to radiation it must be localized in time. In the light 
of these observations we 
consider some possible forms for $\rho_R$. This 
will give us an idea about the possible effects of 
conversion of matter to radiation (or vice versa) on the 
expansion rate of the universe. In fact, 
to have a concrete, 
realistic picture one needs to calculate the cross 
sections and the rates of different such process in 
an expanding universe for different possible scenarios.
We leave 
this ambitious program to detailed studies in future. In 
this subsection we give an idea on the range of possibilities 
by considering some hypothetical choices of $\rho_R$'s 
that are localized in time, and this will be sufficient for our 
purpose. 
First consider the example,
{\it  $\rho_R$=$\sigma\,[1-\tanh^2{\beta(a-a_i)}]$}
where $\sigma$, $\beta$ 
are some constants
and $a_i$ denotes the scale 
factor where $\rho_R$ is maximum.
Then 
$\frac{\dot{\rho}_R}{\rho_R}$=$-2\beta\,
\dot{a}\tanh{\beta(a-a_i)}]$. 
After using      
(\ref{a21d}) one gets
\begin{equation}
\omega_R\,+\,\Delta\omega_R\,=\,
-\frac{\dot{\rho_R}}{3H\rho_R}-1
\,=\,\frac{2}{3}\beta\,a\tanh{\beta(a-a_i)}-1
\label{a5a}
\end{equation}
One notices that
\begin{eqnarray}
&&
\omega_R\,+\,\Delta\omega_R\,
\rightarrow\,-1 ~~~~
\mbox{as}~~ 
a\rightarrow\,0~~\mbox{or}~~ 
a\rightarrow\,a_i\,,
\nonumber \\
&&
\omega_R\,+\,\Delta\omega_R\,
\leq\,-1~~\mbox{when}~~0\,<\,a\,<\,a_i~~
\mbox{and}~~~
\omega_R\,+\,\Delta\omega_R\,
\geq\,-1~~\mbox{when}~~a\,>\,a_i \nonumber
\\
&&\omega_R\rightarrow\,\infty~~~~
\mbox{as}~~ a\rightarrow\,\infty 
,
~~~\mbox{and}~~ 
\omega_R\,+\,\Delta\omega_R\,
\simeq\,0.28~~\mbox{when}~~ 
(a-a_i)\beta\simeq\,2 
\nonumber
\end{eqnarray}
Next consider another example,
{\it $\rho_R$=$\frac{\sigma}{1+\beta(a-a_i)^2}$}. 
Then
$\frac{\dot{\rho}_R}{\rho_R}$=$-\frac{2\beta\,\dot{a}(a-a_i)}
{[1+\beta(a-a_i)^2]^2}$.
After using      
(\ref{a21d}) one gets 
\begin{equation}
\omega_R\,+\,\Delta\omega_R\,
=\,
-\frac{\dot{\rho_R}}{3H\rho_R}-1
\,=\,\frac{2\beta\,a(a-a_i)}{3[1+\beta(a-a_i)^2]}-1
\label{a5b}
\end{equation}
We find that
\begin{eqnarray}
&&
\mbox{For}~~~a=\frac{a_i}{2}~~~~
\omega_R\,+\,\Delta\omega_R\,
=
\,\left(\omega_R\,+\,\Delta\omega_R\right)_{min}
\,=\,-\frac{2\beta\,(\frac{a_i}{2})^2}
{3[1+\beta(\frac{a_i}{2})^2]}-1 \nonumber \\
&&\mbox{and}
~~
\omega_R\,+\,\Delta\omega_R\,>\,
\left(\omega_R\,+\,\Delta\omega_R\right)_{min}
\nonumber \\
&&
\omega_R\,+\,\Delta\omega_R
\rightarrow\,-1 ~~~~
\mbox{as}~~ 
a\rightarrow\,0~~\mbox{or}~~a\rightarrow\,a_i 
~~\mbox{and}
~~~\omega_R\,+\,\Delta\omega_R
\rightarrow\,-\frac{1}{3}~~~~
\mbox{as}~~ a\rightarrow\,\infty
\nonumber
\end{eqnarray}
where the subindex $min$ denotes the minimum value of 
$\omega_R\,+\,\Delta\omega_R$.

In both examples we have 
$\omega_R\,+\,\Delta\omega_R=-1$ 
at $a=a_i$ and 
$\omega_R\,+\,\Delta\omega_R$ 
$\,<\,-1$ for $a\,<\,a_i$, 
$\omega_R\,+\,\Delta\omega_R$ 
$\,>\,-1$ for $a\,>\,a_i$. This is not 
surprising, 
it is just a result of the requiring $\rho_R$ be 
localized. Therefore most of the general lines of the 
above examples are valid in all cases provided that 
$\rho_R$ is localized in time.
Note that both in this set-up and in the standard model 
of cosmology, the universe is radiation dominated at the 
time just after the time of creation of radiation. Moreover 
$\frac{\rho_R}{\rho}\simeq\,1$ when $a\simeq\,a_i$. 
Thus in a model of inflation in this context one may take the 
$\omega_R\,+\,\Delta\omega_R$ values obtained 
for the times close to the time $a=a_i$ 
as a sufficiently realistic 
of the universe for the time just before creation 
of radiation. On the other hand the same is not true for 
other times e.g. for much later times since,
at much later times, the universe almost exhaustively 
is composed of 
components other than radiation e.g. matter (and 
possibly dark energy as well). 
Therefore a more realistic modification of the content of the universe
is needed to extend this scheme to account for early time cosmic
accelerated expansion. 
However regarding dark energy even these simple examples
may give some insight to the problem: A residual 
effect of the $\rho_R$ that survives may have some 
effect for later times. 
Although tiny, $\rho_R$ may 
still have 
a tail for large scale factor, $a(t)$.  
The effect of this tail is determined by how well 
$\rho_R$ is localized, and how big $\rho_R$ and 
$\omega_R\,+\,\Delta\omega_R$ are.
In fact such a model is considered in literature 
\cite{inflation-BEC1}.

\section{Conclusion}

In this paper we have studied, 
as a sub-case of two interacting energy densities,
if the accelerated 
cosmic expansion may be induced by conversion of
extremely non-relativistic particles to radiation.
We have seen that cosmic accelerated expansion
can not be obtained in conversion between matter and radiation
through instantaneous interactions.
In fact thermo dynamical studies give a similar conclusion 
\cite{Scherrer}. 
It seems that the only way to obtain cosmic accelerated expansion
by particle physics interactions is through some intermediate 
state with a negative equation of state
that forms during the conversion. 
It seems difficult to obtain the present
cosmic accelerated expansion wholly through the usual 
particle physics interactions in this way since the localization 
scales of corresponding $\rho_R$'s for the usual 
particle physics processes are at the order of 
atomic scales i.e. 
at scales much smaller than the cosmological scales. 
Even when they have such an effect, these interactions 
will first accelerate the universe 
and then decelerate it in the time 
scale of the interaction 
time (which is smaller than 
$\sim\,10^{-8}\,sec$), hence the net effect would 
be zero. This type of interactions
may be relevant cosmologically only at early times 
(if they involve
the usual particles) 
provided that a significant redshift takes 
place during their interaction time
e.g. during the lifetime of the resonance particle. 
A very early time acceleration may be induced by 
fast out of equilibrium processes 
as those given in Figure \ref{fig:1} 
provided an 
intermediate state with 
$\omega\,<\,0$ forms. Present cosmic accelerated 
expansion may be obtained
in this way only if the life time of the resonance condensate has a
cosmologically 
relevant time scale.
Although we have considered such a toy model in this study, 
in order to entertain these possibilities in detail 
one needs to study different specific models in more detail
along the lines given in this paper and confront it with 
observational data which is beyond the scope of this 
study that aims to seek the degree of 
possibility 
of obtaining the late time and the early time accelerated 
expansions of the universe in this way. Specific models 
along these lines where 
different options for 
$\rho_R$ and $\frac{\tilde{\rho}(t)}{dt}$ 
are specified and their theoretical origins discussed
and whose the results are confronted with 
observational data may be considered in future.

\appendix
\section{Details of the derivations of 
(\ref{a2a}), (\ref{a2c}), (\ref{a2aar}), 
(\ref{a2bar})}

Let at time $t\,<\,t_1$ the energy density of the 
universe, $\rho$ consists of matter (i.e. dust) and 
radiation, 
that is,
\begin{eqnarray}
&&\rho\,=\,\rho_m\,+\,\rho_r\label{1a}\\
&&
\rho_m\,=\,\frac{\rho_{m0}}{[a(t)]^3}~,~~
\rho_r\,=\,\frac{\rho_{r0}}{[a(t)]^4} \label{1b}
\end{eqnarray}

Assume that at $t_1\,<\,t\,<\,t+\Delta\,t$ (where 
$\Delta\,t$ is very small) some energy density 
$(\Delta\tilde{\rho})_1$ is transferred from either of 
the dust 
or the radiation to the other, say, from the dust to the 
radiation e.g. through some decay or other particle 
physics process such as in Figure \ref{fig:1}.
Then the new energy density becomes
\begin{eqnarray}
\rho(t)&=&\{\frac{\rho_{m0}}{[a(t)]^3}
\,-\,(\Delta\tilde{\rho})_1\theta[t-(t_1+\Delta\,t)]\left[
\frac{a(t_1
+\Delta\,t
)}{a(t)}\right]^3\} \nonumber\\
&&+\,\{\frac{\rho_{r0}}{[a(t)]^4} 
\,+\,(\Delta\tilde{\rho})_1\theta[t-(t_1+\Delta\,t)]\left[
\frac{a(t_1+\Delta\,t)}{a(t)}\right]^4\} \nonumber \\
&=&
\frac{\rho_{m0}
\,-\,(\Delta\tilde{\rho})_1\theta[t-(t_1+\Delta\,t)]
[a(t_1+\Delta\,t)]^3}{[a(t)]^3}\nonumber \\
&&+\,
\frac{\rho_{r0}
\,+\,(\Delta\tilde{\rho})_1\theta[t-(t_1+\Delta\,t)]
[a(t_1+\Delta\,t)]^4}
{[a(t)]^4} \label{2}
\end{eqnarray}
where $\theta(t)$ is the Heaviside function (i.e. unit 
step function) with $\theta(t)\,=\,0$ if $t\,<\,0$, and 
$\theta(t)\,=\,1$ if $t\,\geq\,0$. Next assume that at 
$t_1+\Delta\,t\,<\,t\,<\,t_1+2\Delta\,t$ some other 
energy density $(\Delta\tilde{\rho})_2$ is transferred 
from the dust to the radiation, 
and so on. Hence at $t=t_1+n\Delta\,t$ the energy 
density is 
\begin{eqnarray}
&&\rho(t)\,=\,
\frac{\rho_{M\Delta}}{[a(t)]^3}\,+\,
\frac{\rho_{r\Delta}}{[a(t)]^4}\label{3a} 
\end{eqnarray}
where
\begin{eqnarray}
&&\rho_{M\Delta}\,=\,\rho_{m0}
-(\Delta\tilde{\rho})_1\theta[t-(t_1+\Delta\,t)]a^3(t_1+\Delta\,t)
-(\Delta\tilde{\rho})_2\theta[t-(t_1+2\Delta\,t)]
a^3(t_1+2\Delta\,t)\dots \nonumber\\
&&\dots\,-(\Delta\tilde{\rho})_n\theta[t-(t_1+n\Delta\,t)]
a^3[t_1+n\Delta\,t)]
\label{3b} \\
&&
\rho_{r\Delta}\,=\,
\rho_{r0}
+(\Delta\tilde{\rho})_1\theta[t-(t_1+\Delta\,t)]a^4(t_1+\Delta\,t)
+(\Delta\tilde{\rho})_2\theta[t-(t_1+2\Delta\,t)]
a^4(t_1+2\Delta\,t)\dots \nonumber \\
&&\dots\,+
(\Delta\tilde{\rho})_n\theta[t-(t_1+n\Delta\,t)]
a^4[t_1+n\Delta\,t)]
\label{3c}
\end{eqnarray}

As 
$\Delta\,t\rightarrow\,0$, 
$(\Delta\tilde{\rho})_i\rightarrow\,0$ 
($i=1,2,.....$) the energy density in (\ref{3a}) at 
$t\,=\,t_2\,>\,t_1$ becomes
\begin{eqnarray}
&&\rho(t_2)\,=\,\rho_m(t_2)\,+\,\rho_r(t_2) \label{4a}\\
&&
\rho_m(t_2)\,=\,\frac{C_{m}(t_2)}{[a(t_2)]^3}~,~~
\rho_r(t_2)\,=\,\frac{C_{r}(t_2)}{[a(t_2)]^4} \label{4b}
\end{eqnarray}
where 
\begin{eqnarray}
&&C_m(t_2)\,=\,\rho_{m0}
-\,\mbox{lim}_{\;\Delta\,t\rightarrow\,0~,~n\Delta\,t
\rightarrow\,t_2-t_1}\;\{
\left(\frac{(\Delta\tilde{\rho})_1}{\Delta\,t}\right)
\theta[t_2-(t_1+\Delta\,t)]a^3(t_1+\Delta\,t)\nonumber \\
&&+\left(\frac{(\Delta\tilde{\rho})_2}{\Delta\,t}\right)
\theta[t_2-(t_1+2\Delta\,t)]a^3(t_1+2\Delta\,t)\dots\,
+\left(\frac{(\Delta\tilde{\rho})_n}{\Delta\,t}\right)
\theta[t_2-(t_1+n\Delta\,t)]
a^3[t_1+n\Delta\,t)]\} \nonumber \\
&&=\,\rho_{m0}\,-\,\int_{t_1}^{t_2}
\left(\frac{d\tilde{\rho}(t^\prime)}{dt^\prime}\right)
a^3(t^\prime)
\theta(t_2-t^\prime)\,dt^\prime \label{5aa} \\
&&=\,\rho_{m0}\,-\,\int_0^{t_2-t_1}
\left(\frac{d\tilde{\rho}(x)}{dx}\right)_{x=(t_2-u)}
a^3(t_2-u)\theta(u)\,du \label{5ab} \\
&&C_r(t_2)\,=\,
\rho_{r0}
+\,\mbox{lim}_{\;\Delta\,t\rightarrow\,0~,~n\Delta\,t
\rightarrow\,t_2-t_1}\;\{
\left(\frac{(\Delta\tilde{\rho})_1}{\Delta\,t}\right)
\theta[t_2-(t_1+\Delta\,t)]a^4(t_1+\Delta\,t) \nonumber \\
&&+\left(\frac{(\Delta\tilde{\rho})_2}{\Delta\,t}\right)
\theta[t_2-(t_1+2\Delta\,t)]
a^4(t_1+2\Delta\,t)\dots\,
+\left(\frac{(\Delta\tilde{\rho})_n}{\Delta\,t}\right)
\theta[t_2-(t_1+n\Delta\,t)]
a^4[t_1+n\Delta\,t)]\} \nonumber \\
&&=\,\rho_{r0}\,+\,\int_{t_1}^{t_2}
\left(\frac{d\tilde{\rho}(t^\prime)}{dt^\prime}\right)
a^4(t^\prime)\theta(t_2-t^\prime)\,dt^\prime \label{5ba} \\
&&=\,\rho_{r0}\,+\,\int_0^{t_2-t_1}
\left(\frac{d\tilde{\rho}(x)}{dx}\right)_{x=(t_2-u)}a^4(t-u)
\theta(u)\,du \label{5bb}
\end{eqnarray}

From (\ref{4b}) we see that (for $t\,>\,t_1$)
\begin{eqnarray}
&&
\dot{\rho}_m(t)\,=\,-3H\,\rho_m(t)\,+\,
\frac{\dot{C}_{m}(t)}{[a(t)]^3} \label{6a} \\
&&\dot{\rho}_r(t)\,=\,-4H\,\rho_r(t)\,+\,
\frac{\dot{C}_{r}(t)}{[a(t)]^4} \label{6b}
\end{eqnarray}
where over dot denotes derivative with respect to t. 
Eqs. (\ref{6a}) and (\ref{6b}) imply that $\rho_m$ and 
$\rho_r$ satisfy the following equations
\begin{eqnarray}
&&
\dot{\rho}_m(t)\,+\,3H\,\rho_m(t)\,=\,
\frac{\dot{C}_{m}(t)}{[a(t)]^3} \label{7a} \\
&&\dot{\rho}_r(t)\,+\,4H\,\rho_r(t)\,=\,
\frac{\dot{C}_{r}(t)}{[a(t)]^4} \label{7b}
\end{eqnarray}
We see that the energy-momentum tensors for the dust and 
the radiation in this case are conserved. They have 
source terms on the right-hand sides of (\ref{7a}) and 
(\ref{7b}). This is not surprising since there is an 
energy transfer from the dust to the radiation. 
If one 
adds (\ref{7a}) to (\ref{7b}) one obtains  
\begin{equation}
\dot{\rho}_m(t)\,+\,\dot{\rho}_r(t)\,+\,
3H\left[\rho_m(t)\,+\,\frac{4}{3}\rho_r(t)\right]\,=\,
\frac{\dot{C}_{m}(t)}{[a(t)]^3} \,+\,
\frac{\dot{C}_{r}(t)}{[a(t)]^4} \label{8}
\end{equation}
First calculate 
$\dot{C}_m(t)$, that may be found from
\begin{equation}
\dot{C}_m(t)\,=\,\mbox{lim}_{\;\Delta\,t\rightarrow\,0}
\frac{C_m(t+\Delta\,t)\,-\,C_m(t)}{\Delta\,t} 
\label{9}
\end{equation}
In the above formula we employ 
Eqs. (\ref{5ab}) and (\ref{5bb}) (rather 
than Eqs. (\ref{5aa}) and (\ref{5ba})) not to deal with 
variation the $\theta$ function that may lead to 
ambiguity since we do not cover $t^\prime\,>\,t$ in the 
integration. Hence
\begin{eqnarray} 
&&C_m(t+\Delta\,t)\,=\,\rho_{m0}\,-\,
\int_0^{t+\Delta\,t-t_1} 
\left(\frac{d\tilde{\rho}(x)}{dx}\right)_{x=(t+\Delta\,t-u)}
\,a^3(t+\Delta\,t-u)\theta(u)\,du \label{10a} \\
&&C_r(t+\Delta\,t)\,=\,\rho_{r0}\,+\,
\int_0^{t+\Delta\,t-t_1} 
\left(\frac{d\tilde{\rho}(x)}{dx}\right)_{x=(t+\Delta\,t-u)}
\,a^4(t+\Delta\,t-u)\theta(u)\,du \label{10b} 
\end{eqnarray}
Let us calculate these explicitly.
\begin{eqnarray} 
&&\int_0^{t+\Delta\,t-t_1} 
\left(\frac{d\tilde{\rho}(x)}{dx}\right)_{x=(t+\Delta\,t-u)}
\,a^3(t+\Delta\,t-u)\theta(u)\,du \nonumber \\
&=&\int_0^{t-t_1} 
\left(\frac{d\tilde{\rho}}{dx}\right)_{x=(t+\Delta\,t-u)}
\,a^3(t+\Delta\,t-u)\theta(u)\,du \nonumber \\
&&+\,\int_{t-t_1}^{t-t_1+\Delta\,t} 
\left(\frac{d\tilde{\rho}(x)}{dx}\right)_{x=(t+\Delta\,t-u)}
\,a^3(t+\Delta\,t-u)\theta(u)\,du \label{a11}
\end{eqnarray}
For small $\Delta\,t$
\begin{eqnarray}
&&a^3(t+\Delta\,t-u)\,=\,\left[a(t-u)
+\left(\frac{da(x)}{dx}\right)_{x=t-u}\Delta\,t
\,+\,\dots\right]^3\nonumber \\
&&\simeq\,a^3(t-u)+3a^2(t-u)
\left(\frac{da(x)}{dx}\right)_{x=t-u}\Delta\,t 
\label{a12a} \\
&&
\left(\frac{d\tilde{\rho}(x)}{dx}
\right)_{x=t+\Delta\,t-u}\,\simeq\,
\left(\frac{d\tilde{\rho}(x)}{dx}\right)_{x=t-u}\,+\,
\left(\frac{d^2\tilde{\rho}(x)}{dx^2}
\right)_{x=t-u}\Delta\,t\,
\label{a12b}\\
&&\,\int_{t-t_1}^{t-t_1+\Delta\,t} 
\left(\frac{d\tilde{\rho}}{du}\right)(t+\Delta\,t-u)\,
a^3(t+\Delta\,t-u)\theta(u)\,du \nonumber \\
&&\simeq\, 
\left(\frac{d\tilde{\rho}}{dx}\right)_{x=t_1+\Delta\,t}\,
a^3(t_1+\Delta\,t)\theta(t-t_1)\,\Delta\,t \nonumber \\
&&\simeq\,
\left(\frac{d\tilde{\rho}}{dx}\right)_{x=t_1}\,
a^3(t_1)\theta(t-t_1)\,\Delta\,t
\label{a12c}
\end{eqnarray}
Hence we find
\begin{eqnarray}
&&C_m(t+\Delta\,t)\,=\,\rho_{m0}\,-\,
\int_0^{t-t_1} 
\{\left[\left(\frac{d\tilde{\rho}(x)}{dx}\right)_{x=t-u}\,+\,
\left(\frac{d^2\tilde{\rho}(x)}{dx^2}\right)_{x=t-u}
\Delta\,t\right]\nonumber \\
&&\times\,
\left[a^3(t-u)+3a^2(t-u)
\left(\frac{da(x)}{dx}\right)_{x=t-u}\Delta\,t\right]
\,\theta(u)\,du\} \nonumber \\
&-& 
\left(\frac{d\tilde{\rho}(x)}{dx}\right)_{x=t_1}\,
a^3(t_1)\theta(t-t_1)\,\Delta\,t \nonumber \\
&&\simeq\,
C_m(t) \nonumber \\
&&+\,(\Delta\,t)
\int_0^{t-t_1} 
\{3\left(\frac{d\tilde{\rho}(x)}{dx}\right)_{x=t-u}a^2(t-u)
\left(\frac{da(x)}{dx}\right)_{x=t-u} \,+\,
\left(\frac{d^2\tilde{\rho}(x)}{dx^2}\right)_{x=t-u}
a^3(t-u)\}\,\theta(u)\,du \nonumber \\
&+& 
(\Delta\,t) 
\left(\frac{d\tilde{\rho}(x)}{dx}\right)_{x=t_1}\,
a^3(t_1)\theta(t-t_1)\,
\label{a13}
\end{eqnarray}
Note that
\begin{eqnarray}
&&\int_0^{t-t_1} 
\{3\left(\frac{d\tilde{\rho}(x)}{dx}\right)_{x=t-u}a^2(t-u)
\left(\frac{da(x)}{dx}\right)_{x=t-u} \,+\,
\left(\frac{d^2\tilde{\rho}(x)}{dx^2}\right)_{x=t-u}
a^3(t-u)\}\,\theta(u)\,du \nonumber \\
&&=\,\int_0^{t-t_1} 
\{\frac{d}{dx}\left[\left(\frac{d\tilde{\rho}(x)}{dx}\right)a^3(x)
\right]
_{x=t-u}\}\theta(u)\,du \,=\,\int_{t_1}^t 
\{\frac{d}{dx}\left[\left(\frac{d\tilde{\rho}(x)}{dx}\right)a^3(x)
\right]\theta(t-x)\,dx \nonumber \\
&&=\,
-\left(\frac{d\tilde{\rho}(x)}{dx}\right)
a^3(x)|_{x=t_1}^{x=t}\theta(t-t_1)\label{a13a}
\end{eqnarray}

After using (\ref{a13}) and (\ref{a13a}) in (\ref{9}) we find 
\begin{equation}
\dot{C}_m(t)\,=\,
-\left(\frac{d\tilde{\rho}(t)}{dt}\right)
a^3(t)\theta(t-t_1)
\label{14}
\end{equation}
In a similar way one may find $\dot{C}_r(t)$ as
\begin{eqnarray}
&&\dot{C}_r(t)\,=\,
\int_0^{t-t_1} 
\{4\left(\frac{d\tilde{\rho}(x)}{dx}\right)_{x=t-u}a^3(t-u)
\left(\frac{da(x)}{dx}\right)_{x=t-u} \,+\,
\left(\frac{d^2\tilde{\rho}(x)}{dx^2}\right)_{x=t-u}
a^4(t-u)\}\,\theta(u)\,du \nonumber \\
&+& 
\left(\frac{d\tilde{\rho}(x)}{dx}\right)_{x=t_1}\,
a^4(t_1)\theta(t-t_1) 
\nonumber \\
&&=\,
\left(\frac{d\tilde{\rho}(t)}{dt}\right)
a^4(t)\theta(t-t_1)
\label{15}
\end{eqnarray}

\section{An example for the evolution of the energy densities of matter, a resonance with $\omega\,=\,-1$, and radiation as a 
function of scale factor}

In this appendix we derive the evolutions of $C_m$, $\rho_R$, and $C_r$
and the related quantities
in the case
\begin{eqnarray}
&&\omega_R\,=\,-1~~~,~~r\,=\,7~~~~~\mbox{and}\nonumber \\
&&\mbox{Initially}~~s\,=\,-\frac{3}{2}
~,~~~\mbox{in between}~~
s\,=\,0~,~~~\mbox{at late times}~~s\,=\,-2
\label{cn14A}
\end{eqnarray}
We will try to see what happens in each era explicitly. 
First we consider the initial times when 
$t_1\,<\,t\,<\,t_2$, 
$s=-\frac{3}{2}$ i.e. $H=\xi_1\,a^{-\frac{3}{2}}$ 
where $\xi_1$ is some constant. 
Then a cosmological constant-like 
resonance dominated universe where $s=0$ for $t_2\,<\,t\,<\,t_3$, 
Finally, the radiation dominated era at the end where $s=-2$ for $t_3\,<\,t$. For simplicity
We will assume that $\beta$ is constant in all eras. 
We give $C_m$, $\rho_R$, $C_r$ in these eras below by using the formulas 
derived in the text. $\rho_m$ and $\rho_r$ are evident from
$C_m$ and $C_r$. For simplicity we assume abrupt changes in 
the equations of states as one passes from one era to the other although 
in a more realistic case the changes would be smooth.

$C_m$ for $t\,<\,t_2$ may be directly found from (\ref{cn7})
for $r=7$, $s=s_1=-\frac{3}{2}$, $\xi=\xi_1$. The result is
\begin{eqnarray}
&&C_m
\,=\,
\frac{\rho_{m0}}
{1\,+\,
\frac{2\beta\rho_{m0}}{9\xi_1}\theta(t-t_1)\,
\left(a^{r-s-4}\,-\,
a_1^{r-s-4}\right)}
\label{cn7aA}
\end{eqnarray}
where $a=a(t)$, $a_1=a(t_1)$.
$\frac{d\tilde{\rho}_m(t)}{dt}$ in this case
directly follows from
Eq.(\ref{cn12})
\begin{eqnarray}
\left(\frac{d\tilde{\rho}_m(t)}{dt}\right)_1
\,=\,
\frac{81\theta(t-t_1)\,\rho_{m0}^2\beta
\xi_1^2}{4\{-\frac{9}{2}\xi_1\,
\,+\,\beta\,\rho_{m0}\,\theta(t-t_1)\,
\left[a^{\frac{9}{2}}(t_1)\,-\,
a^{\frac{9}{2}}(t)\right]\}^2}
\label{cn16A} 
\end{eqnarray}
After using Eq.(\ref{a21}), 
and Eq.(\ref{cn16A}) we get
\begin{eqnarray}
&&C_R(t)
\,=\,
\frac{81}{4}\,\theta(t-t_1)\,\rho_{m0}^2\beta
\xi_1^{-1}\int_{a_1}^{a}
\,e^{\Gamma(t^\prime-t_1)}
\frac{
a^{\prime\,\frac{1}{2}}da^\prime}
{\{-\frac{9}{2}\xi_1\,
\,+\,\beta\,\rho_{m0}
\left[a_1^{\frac{9}{2}}\,-\,
a^{\prime\,\frac{9}{2}}\right]\}^2}
\label{cn18A}
\end{eqnarray}
where $a\,=\,a(t)$, 
$a_1\,=\,a(t_1)$, and we have used
$dt\,=\,\frac{da}{a\,H}$.
One may let $|\Gamma(t-t_1)|\,\simeq\,0$,
$a_1^{\frac{9}{2}}-a^{\frac{9}{2}}\,\simeq\,0$ since
we consider the times when the production
of $\rho_R$ have just started. Hence (\ref{cn18A}) 
may be approximated by
\begin{eqnarray}
&&C_R(t)
\,=\,
\frac{2}{3}\,\theta(t-t_1)\,\frac{\rho_{m0}^2\beta}
{\xi_1}\left[a^{\frac{3}{2}}\,-\,
a_1^\frac{3}{2}\right]
\label{cn18aA}
\end{eqnarray}
$\rho_R(t)$, by (\ref{a21a}), becomes 
\begin{eqnarray}
&&\rho_R(t)\,=\,
\,e^{-\Gamma(t-t_1)}\frac{C_R}{a^3}
\,\simeq\,
\frac{2}{3}\,\theta(t-t_1)\,\frac{\rho_{m0}^2\beta}
{\xi_1}\frac{1}{a^3(t)}
\left[a^{\frac{3}{2}}\,-\,a_1^\frac{3}{2}\right]
\label{cn19A} \\
&&~~~\mbox{where}~~~~t_1\,<\,t\,<\,t_2 \nonumber \\
\end{eqnarray}
The corresponding $C_r$ is found by using 
(\ref{a2cr1}) and (\ref{cn19A}) as
\begin{equation}
C_r(t)\,=\,
\frac{2\Gamma\rho_{m0}^2\beta}{\xi_1^2}
\int_{a_1}^a\,(a^{\prime\,3}\,-\,
a_1^\frac{3}{2}a^{\prime\,\frac{3}{2}})\,da^\prime
\,=\, 
\frac{\Gamma\rho_{m0}^2\beta}{2\xi_1^2}\left[a^4\,-\,
\frac{8a_1^\frac{3}{2}}{5}a^\frac{5}{2}\,+\,\frac{3}{5}a_1^4\right]
\label{cn19crA}
\end{equation}

At later times we assume the condensate of the resonance dominates, 
so $s=0$ i.e. $H=\xi_2$. To determine $\rho_R$ we should first find $C_m$ in
this case. To find $C_m$ in this case we should divide
the integral in (\ref{cn6}) into two parts; one for the era when
$s=-\frac{3}{2}$ between $t_1$ and $t_2$, and then the era when
$s=0$ between $t_2$ and $t$. Then the corresponding
$C_m(t)$ is found as  
\begin{eqnarray}
-&&\frac{1}{C_m(t)}
\,=\,
-\theta(t-t_1)\,\beta\,\left[
\frac{1}{\xi_1}\int_{a_1}^{a_2}\frac{\,da^\prime}{a^{\prime\,5-r_1+s_1}}
\,+\,\frac{1}{\xi_2}\int_{a_2}^{a}\frac{\,da^\prime}{a^{\prime\,5-r_2+s_2}}
\right]
-\frac{1}{C_m(t_1)} \nonumber \\
&&=\,
\beta\,\theta(t-t_1)\{
\frac{1}
{\xi(4-r_1+s_1)}\left[
\frac{1}{a^{4-r_1+s_1}(t_2)} \,-\,
\frac{1}{a^{4-r_1+s_1}(t_1)}\right]\nonumber \\
&&+\,
\frac{1}
{\xi(4-r_2+s_2)}\left[
\frac{1}{a^{4-r_2+s_2}(t)} \,-\,
\frac{1}{a^{4-r_2+s_2}(t_2)}\right]\,-\,
\frac{1}{\rho_{m0}} \nonumber \\
&&=\,
\{\rho_{m0}\beta[
\theta(t_2-t_1)\xi_2(4-r_2+s_2)
\left(a_1^{4-r_1+s_1} \,-\,
a_2^{4-r_1+s_1}\right)a_2^{4-r_2+s_2}a^{4-r_2+s_2}
\nonumber \\ 
&&+\,\theta(t-t_2)\xi_1(4-r_1+s_1)
\left(a_2^{4-r_2+s_2} \,-\,
a^{4-r_2+s_2}\right)
a_2^{4-r_1+s_1}a_1^{4-r_1+s_1}
] \nonumber \\
&&-\,\xi_1\xi_2(4-r_1+s_1)(4-r_2+s_2) 
a_2^{4-r_1+s_1}a_1^{4-r_1+s_1}
a_2^{4-r_2+s_2}a^{4-r_2+s_2}\} \nonumber \\
&&\times\,\left[\rho_{m0}\xi_1\xi_2
(4-r_1+s_1)(4-r_2+s_2)
a_2^{4-r_1+s_1}a_1^{4-r_1+s_1}
a_2^{4-r_2+s_2}a^{4-r_2+s_2}\right]^{-1}
\label{cn19aaA}
\end{eqnarray}
Hence we find $C_m$ in this case as
\begin{eqnarray}
C_m(t)&=&\rho_{m0}\{
1\,-\,
\rho_{m0}\beta[
\theta(t_2-t_1)\xi_1^{-1}(4-r_1+s_1)^{-1}
(a_2^{r_1-4-s_1}\,-\,
a^{r_1-4-s_1})
 \nonumber \\
&&+\,\theta(t-t_2)\xi_2^{-1}(4-r_2+s_2)^{-1}
(a^{r_2-4-s_2} \,-\,
a_2^{r_2-4-s_2})]\}^{-1}
\label{cn19bA}
\end{eqnarray}
Thus, the corresponding 
$\frac{d\tilde{\rho}_m(t)}{dt}$ 
for $r_1=r_2=7$, $s_1=-\frac{3}{2}$, $s_2=0$, and
$t\,>\,t_2$ is
\begin{eqnarray}
&&\left(\frac{d\tilde{\rho}_m(t)}{dt}\right)_2
\,=\,-\frac{\dot{C}_m(t)}{a^3(t)} \nonumber \\
&&=\,
\frac{\theta(t-t_2)\,\rho_{m0}^2\beta}
{\{1
\,+\,\frac{1}{3}
\beta\,\rho_{m0}\,\left[
\theta(t_2-t_1)\,
\frac{2}{3}\xi_1^{-1}(a_2^{\frac{9}{2}}\,-\,
a_1^{\frac{9}{2}})
\,+\,
\theta(t-t_2)\,\xi_2^{-1}(a^3\,-\,
a_2^3)\right]\}^2}
\label{cn20aA}
\end{eqnarray}
where we have skipped the term proportional to
the delta function $\delta(t-t_2)$ since it does not
contribute to $C_R$ below. 
\begin{eqnarray}
&&C_R(t)
\,=\,
\int_{t_1}^{t_2}dt^\prime
\left(\frac{d\tilde{\rho}_m(t^\prime)}{dt^\prime}
\right)_1
\theta(t-t^\prime)
\,e^{\Gamma(t^\prime-t_1)}\,+\,
\int_{t_2}^{t}dt^\prime
\left(\frac{d\tilde{\rho}_m(t^\prime)}{dt^\prime}
\right)_2
\theta(t-t^\prime)
\,e^{\Gamma(t^\prime-t_1)}\nonumber \\
&&\simeq\,
\frac{2}{3}\,\frac{\rho_{m0}^2\beta}
{\xi_1}\left[a_2^{\frac{3}{2}}\,-\,
a_1^\frac{3}{2}\right]
\,+\,\theta(t-t_2)\,
\rho_{m0}^2\beta\xi_2^{-1}
\int_{a_2}^{a}
\,
\frac{
\left(\frac{a^\prime}{a_1}\right)^\gamma
da^\prime}
{a^\prime\{A
\,+\,B
\theta(t-t_2)
a^{\prime\,3}\}^2}
\label{cn20bA}
\end{eqnarray}
where $\gamma=\frac{\Gamma}{H}$,
$a(t)$=$e^{Ht}$, and
\begin{eqnarray}
&&A\,=\,
1\,+\,
\frac{1}{3}\beta\,\rho_{m0}\,
\left[
\frac{2}{3}\xi_1^{-1}(a_2^{\frac{9}{2}}\,-\,
a_1^{\frac{9}{2}})
\,-\,
\xi_2^{-1}a_2^3 \right]
\label{cn20aaA} \\
&&B\,=\,
\frac{1}{3}\xi_2^{-1}
\beta\,\rho_{m0} \label{cn20abA} 
\end{eqnarray}
In general the result of
(\ref{cn20bA}) is complicated, so analyzing 
the result is not easy to see. However one may
have an idea on the form of $\rho_R$ by analyzing some
specific cases. One must have $\Gamma\,>\,H$ to insure 
that the system stays coupled, so the decay may take place.
One must also take $\Gamma$ in the same order of magnitude 
as the Hubble parameter $H$ to insure that the decay process
is relevant at cosmological scales.  
In the light of these observations we may, for example, 
let $\gamma=3$. Hence
(\ref{cn20bA}) becomes
\begin{eqnarray}
&&C_R(t)
\,=\,
\frac{2}{3}\,\frac{\rho_{m0}^2\beta}
{\xi_1}\left[a_2^{\frac{3}{2}}\,-\,
a_1^\frac{3}{2}\right]
\,+\,
\theta(t-t_2)\,
\frac{\rho_{m0}^2\beta}{3B\xi_2}
\left[\frac{1}{A}\,-\,
\frac{1}
{A\,+\,
B\,a^3}\right]
\label{cn20cA}
\end{eqnarray}
It is evident from (\ref{cn20cA}) that $C_R$ approaches
the constant value $\rho_{m0}a_1^{-3}$ and its rate of increase decreases as the scale factor 
$a$ increases, as expected. 
$\rho_R(t)$, by (\ref{a21a}), becomes 
\begin{eqnarray}
&&\rho_R(t)\,=\,
\left(\frac{a_1}{a}\right)^3\frac{1}{3}
\,\left[\frac{2\rho_{m0}^2\beta}
{\xi_1}\{a_2^{\frac{3}{2}}\,-\,
a_1^\frac{3}{2}\}
\,+\,
\theta(t-t_2)\,
\frac{\rho_{m0}^2\beta}{B\xi_2}\{\frac{1}{A}\,-\,
\frac{1}{A\,+\,
B\,a^3}\}\right]
\label{cn20dA} \\
&&~~~\mbox{where}~~~~t_2\,<\,t\,<\,t_3 \nonumber 
\end{eqnarray}
here the $\left(\frac{a_1}{a}\right)^3$ term in the front of
(\ref{cn20dA}) is due to the $e^{-\Gamma(t-t_1)}$ term in (\ref{a21a}). 
The corresponding $C_r$ is found as
\begin{eqnarray}
C_r(t)&=&
\Gamma\,\left[
\int_{t_1}^{t_2}
\rho_R(t^\prime)
a^4(t^\prime)
\theta(t-t^\prime)\,dt^\prime 
\,+\,
\int_{t_2}^{t}
\rho_R(t^\prime)
a^4(t^\prime)
\theta(t-t^\prime)\,dt^\prime \right] \nonumber\\
&&=\,
\frac{\Gamma\rho_{m0}^2\beta}{2\xi_1^2}\left[a_2^4\,-\,
\frac{8a_1^\frac{3}{2}}{5}a_2^\frac{5}{2}
\,+\,\frac{3}{5}a_1^4\right] \nonumber\\
&&+\,
\frac{\Gamma\rho_{m0}^2\beta\,a_1^3}{3\xi_2}
\int_{a_2}^a
\,\left[\frac{2}{\xi_1}\{a_2^{\frac{3}{2}}\,-\,
a_1^\frac{3}{2}\}
\,+\,
\frac{1}{B\xi_2}\{\frac{1}{A}\,-\,
\frac{1}{A\,\,+\,
B\,a^{\prime\,3}}\}\right]\,da^\prime \nonumber\\
&&=\,\frac{\Gamma\rho_{m0}^2\beta}{2\xi_1^2}\left[a_2^4\,-\,
\frac{8a_1^\frac{3}{2}}{5}
a_2^\frac{5}{2}
\,+\,\frac{3}{5}a_1^4\right]
\,+\,
\frac{2\Gamma\rho_{m0}^2\beta\,a_1^3}{3\xi_1\xi_2}
(a_2^\frac{3}{2}\,-\,
a_1^\frac{3}{2})(a\,-\,a_2) \nonumber \\
&&\,+\,
\frac{\Gamma\rho_{m0}^2\beta\,a_1^3}{3B\xi_2^2}
[\frac{a-a_2}{A}\,+\,(6A^\frac{2}{3}B^\frac{1}{3})^{-1}
\{ 2\sqrt{3}\,tan^{-1}\left(\frac{1-\frac{2B^\frac{1}{3}}{A^\frac{1}{3}}a}{\sqrt{3}}\right)
\,-\,
2\sqrt{3}\,tan^{-1}\left(\frac{1-\frac{2B^\frac{1}{3}}{A^\frac{1}{3}}a_2}{\sqrt{3}}\right)
\nonumber \\
&&\,-\,2\ln{\left(\frac{A^\frac{1}{3}+B^\frac{1}{3}\,a}
{A^\frac{1}{3}+B^\frac{1}{3}\,a_2}\right)}
\,+\,\ln{\left(\frac{A^\frac{2}{3}
-A^\frac{1}{3}B^\frac{1}{3}
\,a
+B^\frac{2}{3}\,a^2}{A^\frac{2}{3}
-A^\frac{1}{3}B^\frac{1}{3}
\,a_2
+B^\frac{2}{3}\,a_2^2}\right)}\}]
\label{cn20crA}
\end{eqnarray}
The rate of change 
$C_r$ in (\ref{cn20crA}), and if it increases for reasonable values of parameters 
is not evident from(\ref{cn20crA}) since the form
of the expression is rather complicated. Therefore one may check the rate of change of 
$C_r$ in (\ref{cn20crA}) for a few values of phenomenologically viable sets of parameters. The results of the corresponding
$C_r$ versus $a$ plots suggest that $C_r$ increases as $\propto\,a$ with the scale factor 
in this era, $t_1\,<\,t\,<\,t_2$. 
For example, for 
$\rho_{m0}\simeq\,\rho_c\,=\,9\times\,10^{-30}\,g\,cm^{-3}$,
$\beta^\prime\,=\,\frac{1}{4}$, $\sigma(t_1)\,=\,10^{-27}\,cm^2$, 
$v(t_1)\,=\,10^5\,cm/s$, $n_0\,=\,4\times\,10^{-5}\,cm^{-3}$ (where 
$\beta^\prime$,$\sigma(t_1)$, $v(t_1)$ are defined in (\ref{c1}) and 
$n_0$ is the number density of matter particles at time $t_1$) gives
$\xi_1$=$\sqrt{\frac{8\pi\,G}{3}\rho_{m0}}\,\simeq\,
2.2\times\,10^{-3}\,s^{-1}$,
$\beta\rho_{m0}$=$10^{-27}\,s^{-1}$. Moreover one may relate $\xi_2$ to $\xi_1$ by
$\xi_1\,=\,a_1^\frac{3}{2}\xi_2$ where we have used the fact that at time $t_1$ the total energy
density in a unit comoving volume $a^3$ lost by the matter i.e. 
$\rho_{m0}=\frac{3\xi_1^2}{8\pi\,G}$ is equal to the one obtained by the resonance i.e. to
$\frac{3\xi_3^2a^3}{8\pi\,G}$. 
$\xi_1\,=\,a_1^\frac{3}{2}\xi_2$ implies that $\xi_2\,>\,2.2\times\,10^{-3}\,s^{-1}$. Therefore $A\,\simeq\,1$,
$B\,\simeq\,0$, and from the definition of $B$ we have $\xi_2\,B=\frac{1}{3}\beta\rho_{m0}$. 
In this approximation the last two lines in (\ref{cn20crA}) cancel after expanding 
$tan^{-1}$ and $\ln{}$ by Taylor expansion and 
keeping the leading order terms and then letting $B=0$, $A=1$.
Thus (\ref{cn20crA}) may be 
approximated by
\begin{equation}
C_r(t)\,\simeq\,
\frac{\Gamma\rho_{m0}}{\xi_1}
\left[\frac{\rho_{m0}\beta}{2\xi_1}
\left(a_2^4-
\frac{8a_1^\frac{3}{2}}{5}
a_2^\frac{5}{2}
+\frac{3}{5}a_1^4\right)
\,+\,
\left(\frac{2\rho_{m0}\beta\,a_1^3(a_2^\frac{3}{2}-
a_1^\frac{3}{2})
}{3\xi_2}
\right)
(a-a_2)\right]
\label{cn21crA}
\end{equation}
We see that $C_r$ continues to increase in the era, $t_1\,<\,t\,<\,t_2$
while its rate of increase gets smaller as expected.

\begin{figure}
\includegraphics[scale=0.9]{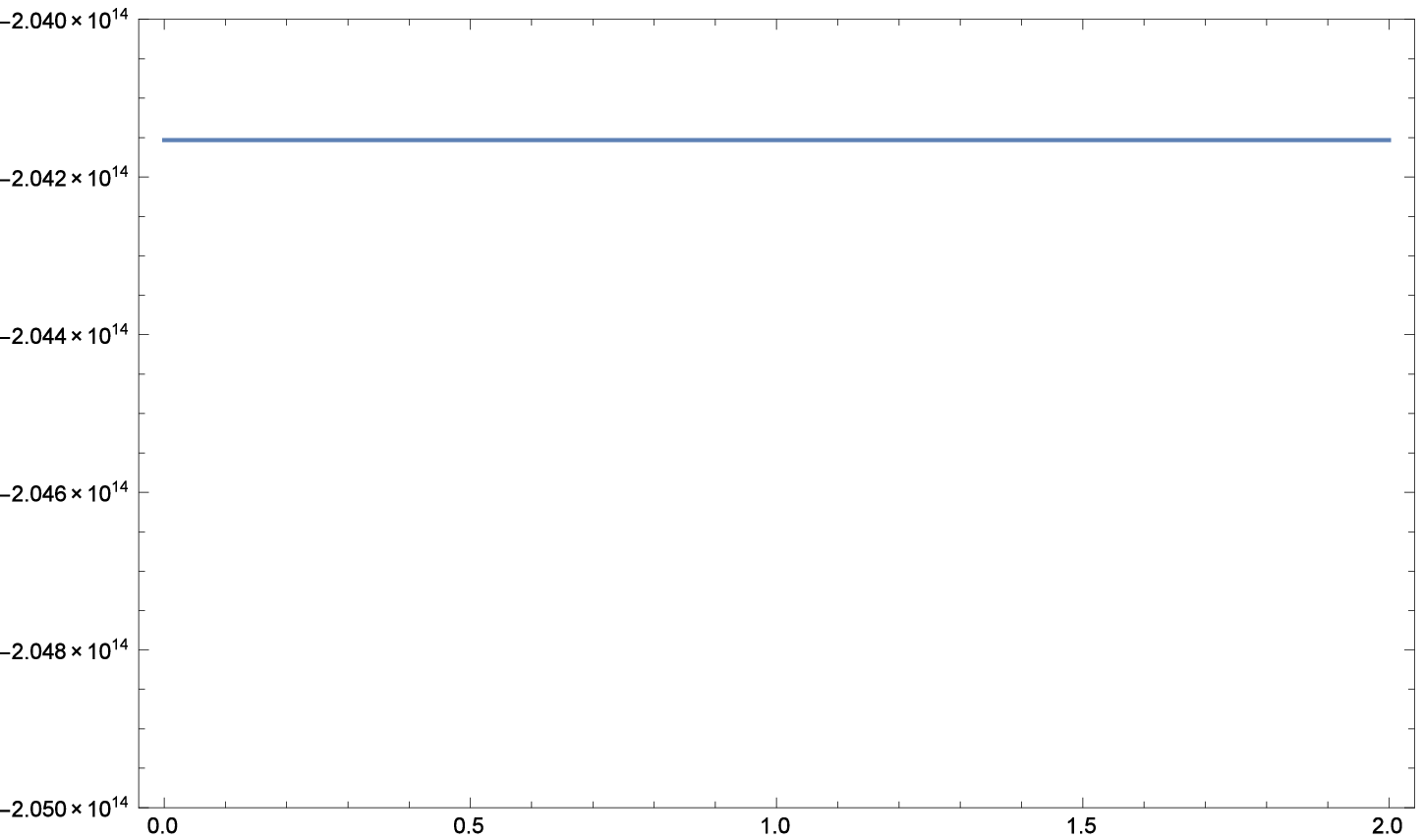}
\caption{The value of the integral in Equation (\ref{cn25A})
as a function of scale factor for $A^\prime\,=\,1$,
$B^\prime\,=\,10^{-20}$}
\label{fig:2}
\end{figure}

\begin{figure}[h]
\includegraphics[scale=0.9]{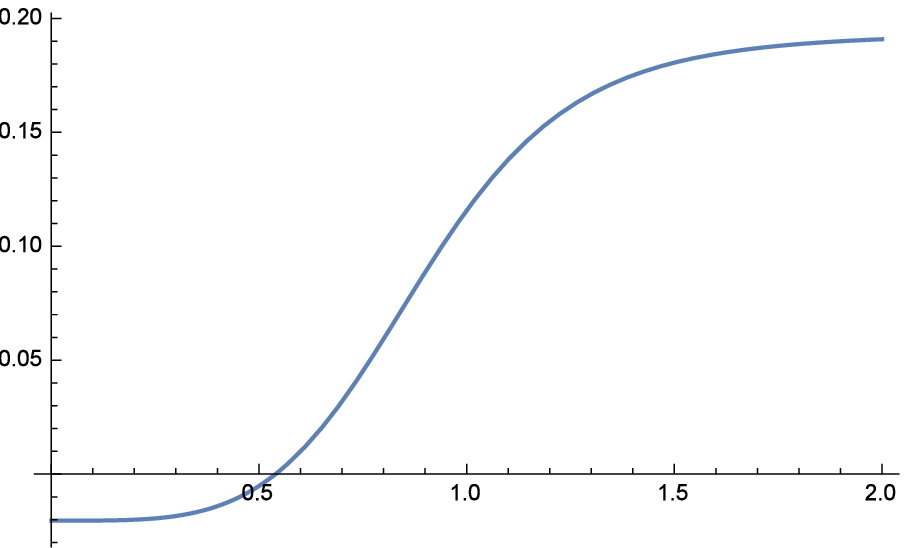}
\caption{The value of the integral in Equation (\ref{cn25A})
as a function of scale factor for $A^\prime\,=\,1$,
$B^\prime\,=\,1$}
\label{fig:3}
\end{figure}

\begin{figure}[h]
\includegraphics[scale=0.9]{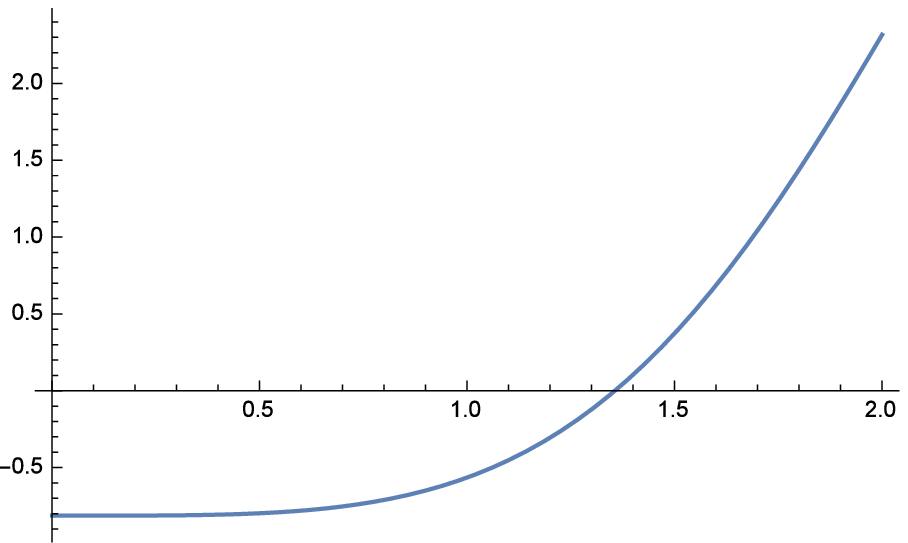}
\caption{The value of the integral in Equation (\ref{cn25A})
as a function of scale factor for $A^\prime\,=\,1$,
$B^\prime\,=\,0.01$}
\label{fig:4}
\end{figure}

Finally we consider the last era of the conversion of matter to radiation 
through formation of a resonance i.e. the radiation dominated 
era after conversion of most of matter to radiation i.e. when $s=-2$.  
We find $C_m^{-1}$ in this case in a similar way as done in (\ref{cn19aaA}),
that is,
\begin{eqnarray}
\frac{1}{C_m(t)}
&=&
\frac{\beta}{\xi_1}\int_{a_1}^{a_2}\frac{\,da^\prime}{a^{\prime\,5-r_1+s_1}}
\,+\,\frac{\beta}{\xi_2}\int_{a_2}^{a_3}\frac{\,da^\prime}{a^{\prime\,5-r_2+s_2}}
\,+\,\frac{1}{C_m(t_1)} \nonumber \\
&&+\,
\frac{\theta(t-t_3)\,
\beta}{\xi_3}\int_{a_3}^{a}\frac{\,da^\prime}{a^{\prime\,5-r_1+s_1}}
\label{cn21aA}
\end{eqnarray}
Then $C_m$ for $t\,>\,t_3$ is found as 
\begin{equation}
C_m(t)\,=\,\frac{\rho_{m0}}{A^\prime\,+\,\theta(t-t_3)\,
B^\prime\,a^5}
\label{cn21bA}
\end{equation}
where $A^\prime$, $B^\prime$ are some constants that, for
$r_1=r_2=r_3=7$, $s_1=-\frac{3}{2}$,
$s_2=0$, $s_1=-2$,  given by
\begin{equation}
A^\prime\,=\,
1\,+\,
\frac{1}{3}\beta\rho_{m0}
\left[
\frac{2}{3}\xi_1^{-1}\left(
a_2^{\frac{9}{2}}-\
a_2^{\frac{9}{2}}\right)\,+\,
\xi_2^{-1}\left(a_3^3-a_2^3\right)\right]\,-\,
\frac{a_3^5\theta(t-t_3)}{5\xi_3}~,~~
B^\prime\,=\,
\frac{1}{5}\xi_3^{-1}
\beta\,\rho_{m0} \label{cn21bbA} 
\end{equation}
The corresponding 
$\frac{d\tilde{\rho}_m(t)}{dt}$ is
\begin{eqnarray}
\left(\frac{d\tilde{\rho}_m(t)}{dt}\right)_3
\,=\,-\frac{\dot{C}_m(t)}{a^3(t)} \,=\,
\theta(t-t_3)\,
\frac{\rho_{m0}B\,H\,a^2}{(A\,+\,B\,a^5)^2}
\label{cn22A}
\end{eqnarray}
where we have skipped the term proportional to
the delta function $\delta(t-t_2)$ since it does not
contribute to $C_R$ below. 
Thus $C_R$ is found as
\begin{eqnarray}
&&C_R(t)
\,=\,
\frac{2}{3}\,\frac{\rho_{m0}^2\beta}
{\xi_1}\left[a_2^{\frac{3}{2}}\,-\,
a_1^\frac{3}{2}\right]
\,+\,
\theta(t-t_2)\,
\frac{1}{3B\xi_2}\rho_{m0}^2\beta\left[\frac{1}{A}\,-\,
\frac{1}{A\,\,+\,B\,a_3^3}\right]\nonumber \\
&&+\,B\rho_{m0}
\theta(t-t_3)\,
\int_{a_3}^a
\,e^{\frac{\Gamma}{2\xi_3}\left(a^{\prime\,2}-a_1^2\right)}
\frac{a^\prime\,da^\prime}{\{A^\prime
\,+\,B^\prime\,a^{\prime\,5}\}^2}
\label{cn23A}
\end{eqnarray}
where we have used the fact that $H=\xi_3a^{-2}$ implies 
$t-t_1$=$\frac{1}{2\xi_3}\left(a^2-a_1^2\right)$. The integral in
(\ref{cn23A}) can not be evaluated exactly even with use of Mathematica.
However one may give an approximate result as follows. At the final stages of the resonance
we have $\Gamma(t-t_1)\,\sim\,1$, so $x\,=\,1\,-\,\Gamma(t-t_1)\,<<\,1$. Therefore 
\begin{equation}
e^{\Gamma(t-t_1)}\,=\,
e^{1-x}\,=\,e\,e^{-x}\,
\,\simeq\,e\,(1-x)\,=\,e\,
\Gamma(t-t_1)
\label{cn24A}
\end{equation}
Hence the integral in (\ref{cn23A}) may be approximated as
\begin{eqnarray}
&&\int_{a_3}^a
\,e^{\frac{\Gamma}{2\xi_3}\left(a^{\prime\,2}-a_1^2\right)}
\frac{a^\prime\,da^\prime}{\{A^\prime
\,+\,B^\prime\,a^{\prime\,5}\}^2} \nonumber \\
&&\simeq\,
\int_{a_3}^{a}
\,\frac{e\,\Gamma}{2\xi_3}
\frac{\left(a^{\prime\,2}-a_1^2\right)
a^\prime\,da^\prime}{\{A^\prime\,+\,B^\prime\,a^{\prime\,5}\}^2}
\,\simeq\,
\frac{e\,\Gamma}{2\xi_3}
\int_{a_3}^{a}
\,\frac{a^{\prime\,3}\,da^\prime}
{\{A^\prime\,+\,B^\prime\,a^{\prime\,5}\}^2}
\label{cn25A}
\end{eqnarray}
The result of (\ref{cn25A}) is rather complicated so that
it makes it difficult to get concrete results from the expression.
However one may draw its plots for several values of $A^\prime$ and
$B^\prime$. The indefinite form of (\ref{cn25A}) for all tried values of
$A^\prime$, $B^\prime$ give flat graphs for almost all values of $a(t)$.
For example, for reasonable values of parameters discussed 
after Eq. (\ref{cn20crA}) and using the conservation of energy in a 
comoving volume at time $t_2$ i.e.
$\xi_2\,a_2^\frac{3}{2}\,=\,\xi_3\,a_2^{-\frac{1}{2}}$ i.e.  
$\xi_3\,=\,\xi_2\,a_2^2\,=\,\xi_1\,a_1^{-\frac{3}{2}}\,a_2^2$ if the resonance 
lives a sufficiently long time so that $a_2\,>>\,a_1$ then we have 
$\xi_3\,<\,\xi_1\,<\,\xi_2$. Otherwise $\xi_1$ and $\xi_3$ must be comparable.
Therefore $A^\prime\simeq\,1$, $B^\prime\simeq\,0$. 
We have given the result of this integration for two three different ratios of $A^\prime$ 
and $B^\prime$ in the figures 
Fig.\ref{fig:2}, 
Fig.\ref{fig:3}, 
Fig.\ref{fig:4}. 
Therefore the result of (\ref{cn25A}) is almost zero for all
phenomenologically viable cases. In other words in this final era i.e. for 
$t\,>\,t_3$
$C_R$ does not
increase any more and $\rho_R$ decays exponentially
\begin{eqnarray}
&&\rho_R(t)
\,=\,e^{\Gamma(t-t_1)}\,C_R(t)\nonumber \\
&&\simeq\,
\frac{4\xi_3\rho_{m0}^2\beta
}{3e\,\Gamma\xi_1}
\left(a^2-a_1^2\right)^{-1}
\left[a_2^\frac{3}{2}\,-\,
a_1^\frac{3}{2}\,
\,+\,
\theta(t-t_2)\,
\frac{a_1^\frac{3}{2}}{2B}
\left(1\,-\,
\frac{1}{1\,\,+\,B\,a_3^3}\right)\right]
\label{cn26A}
\end{eqnarray}
 Then the corresponding $C_r$ becomes
\begin{eqnarray}
&&C_r(t)\,\simeq\,
\frac{\Gamma\rho_{m0}}{\xi_1}
\left[\frac{\rho_{m0}\beta}{2\xi_1}
\left(a_2^4-
\frac{8a_1^\frac{3}{2}}{5}
a_2^\frac{5}{2}
+\frac{3}{5}a_1^4\right)
\,+\,
\left(\frac{2\rho_{m0}\beta\,a_1^3(a_2^\frac{3}{2}-
a_1^\frac{3}{2})
}{3\xi_2}
\,+\,
a_1^\frac{9}{2}\right)
(a_3-a_2)\right] \nonumber \\
&&+\,C^\prime\,\int_{a_3}^a\,\frac{a^{\prime\,5}\,da^\prime}
{a^{\prime\,2}-a_1^2}
\label{cn26crA}
\end{eqnarray}
where
\begin{equation}
C^\prime\,=\,
\frac{2\rho_{m0}^2\beta
}{3e\,\xi_1}
\left[
2\left(a_2^\frac{3}{2}\,-\,
a_1^\frac{3}{2}\right)\,
\,+\,
\frac{a_1^\frac{3}{2}}{B}
\left(1\,-\,
\frac{1}{1\,\,+\,B\,a_3^3}\right)\right]
~~~\mbox{for}~~t\,>\,t_3
\label{cn26craA}
\end{equation}
After integration of the integral in (\ref{cn26crA}) we obtain
\begin{eqnarray}
&&C_r(t)\,\simeq\,
\frac{\Gamma\rho_{m0}}{\xi_1}
\left[\frac{\rho_{m0}\beta}{2\xi_1}
\left(a_2^4-
\frac{8a_1^\frac{3}{2}}{5}
a_2^\frac{5}{2}
+\frac{3}{5}a_1^4\right)
\,+\,
\left(\frac{2\rho_{m0}\beta\,a_1^3(a_2^\frac{3}{2}-
a_1^\frac{3}{2})
}{3\xi_2}
\,+\,
a_1^\frac{9}{2}\right)
(a_3-a_2)\right] \nonumber \\
&&+\,C^\prime\,\left[\frac{a_1^2}{2}\left(a^2\,-\,a_3^2\right)
\,+\,\frac{a^4\,-\,a_3^4}{4}\,+\,
\frac{a_1^4}{2}\,\ln{\left(\frac{
a^2\,-\,a_1^2}{a_3^2\,-\,a_1^2}\right)}\right]
\label{cn26crbA}
\end{eqnarray}
We notice that there is a term proportional to $a^4$.
Although the coefficient of this much smaller than the others 
it is anomalous since it  makes the production of radiation seem gets faster
while it should be go to a constant value at the end of the decay. In fact 
this anomalous behavior is due to identifying the time when 
$\Gamma(t-t_1)\simeq\,1$ with radiation dominated era. Taking this time as 
a radiation dominated era is an incorrect identification since about one third 
of resonance particles with $s=0$ survive at this time and the increase in $C_r$ 
due to decay of resonance makes it effectively behave as an energy density with 
equation of state smaller than $\frac{1}{3}$. Therefore taking $s=-2$ is not valid
unless a small amount of resonance particles survive and hence the rate of 
conversion of resonance particles to radiation becomes very small. Therefore one must 
take $s=-2$ at a much later time. For example one may take   
$\Gamma(t-t_1)\simeq\,4$  i.e. 
\begin{eqnarray}
&&\frac{\Gamma(t-t_1)}{4}\,\sim\,1~~~\Rightarrow~~
x=1-\frac{\Gamma(t-t_1)}{4}\,<<\,1~~~~\Rightarrow~~
e^{\Gamma(t-t_1)}\,=\,e^{4(1-x)}\,=\,
e^4\left(e^{-x}\right)^4\, \nonumber \\
&&\simeq\,
e^4\left(1-x\right)^4
\,=\,e^4\left(\frac{\Gamma(t-t_1)}{4}\right)^4
\,=\,\left(\frac{e\Gamma}{8\xi_3}\right)^4(a^2-a_1^2)^4
\label{cn27A}
\end{eqnarray}
After substituting (\ref{cn27A}) in the first term in (\ref{cn25A}) and using Mathematica to evaluate
the integral and draw the corresponding plots we get almost zero for the result of the
integration as before for reasonable values of parameters mentioned before. 
Next we use (\ref{cn27A}) to find $\rho_R$. Then  
(\ref{cn26A}) is replaced by
\begin{eqnarray}
\rho_R(t)
&=&e^{\Gamma(t-t_1)}\,C_R(t)
\simeq\,
\frac{4096\xi_3^4\rho_{m0}^2\beta}{3e^4\,\Gamma^4\xi_1}
\left(a^2-a_1^2\right)^{-4} \nonumber \\
&&\times\,
\left[2\left(a_2^\frac{3}{2}\,-\,
a_1^\frac{3}{2}\right)\,
\,+\,\theta(t-t_2)\,
\frac{a_1^\frac{3}{2}}{B}
\left(1\,-\,
\frac{1}{1\,+\,B\,a_3^3}\right)\right]
~~~\mbox{for}~~t\,>\,t_3
\label{cn28A}
\end{eqnarray}
Then the corresponding $C_r$ becomes
\begin{eqnarray}
&&C_r(t)\,\simeq\,
\frac{\Gamma\rho_{m0}}{\xi_1}
\left[\frac{\rho_{m0}\beta}{2\xi_1}
\left(a_2^4-
\frac{8a_1^\frac{3}{2}}{5}
a_2^\frac{5}{2}
+\frac{3}{5}a_1^4\right)
\,+\,
\left(\frac{2\rho_{m0}\beta\,a_1^3(a_2^\frac{3}{2}-
a_1^\frac{3}{2})
}{3\xi_2}
\,+\,
a_1^\frac{9}{2}\right)
(a_3-a_2)\right] \nonumber \\
&&+\,C^{\prime\prime}\,
\int_{a_3}^a\,\frac{a^{\prime\,5}\,da^\prime}
{\left(a^{\prime\,2}-a_1^2\right)^4}
\label{cn28crA}
\end{eqnarray}
where
\begin{equation}
C^{\prime\prime}\,=\,
\frac{4096\xi_3^3\rho_{m0}^2\beta}{3e^4\,\Gamma^3\xi_1}
\left[
2\left(a_2^\frac{3}{2}\,-\,
a_1^\frac{3}{2}\right)\,
\,+\,
\frac{a_1^\frac{3}{2}}{B}
\left(1\,-\,
\frac{1}{1\,\,+\,B\,a_3^3}\right)\right]
~~~\mbox{for}~~t\,>\,t_3
\label{cn28craA}
\end{equation}
The result of the integral in (\ref{cn28crA}) is
\begin{equation}
\int_{a_3}^a\,\frac{a^{\prime\,5}\,da^\prime}
{\left(a^{\prime\,2}-a_1^2\right)^4}\,=\,
\frac{a_1^2\,-\,3a_1^2a^2\,+\,3a^4}{6(a^2-a_1^2)}\,-\,
\frac{a_1^2\,-\,3a_1^2a_3^2\,+\,3a_3^4}{6(a_3^2-a_1^2)}
\label{cn28crbA}
\end{equation}
which goes to a constant value (i.e. $C_r$ goes to a constant value) 
as $a$ increases as expected. One may check that later times 
(that may be taken as the time that satisfy $\Gamma(t-t_1)\simeq\,k$ where $k\,>\,4$) 
$C_r$ goes to the constant value 
even faster.

\begin{acknowledgments}
I would like to thank Professor Luca Amendola for pointing out a crucial calculational error
in the draft version of this paper. I would also like to thank Professor Masahiro Morikawa
for his valuable detailed comments regarding their papers related to the use of Bose-Einstein condensation
for cosmic accelerated expansion. 
\end{acknowledgments}


\end{document}